\documentclass[aps,a4paper,floatfix,nofootinbib]{revtex4}

\usepackage{graphicx,color}
\usepackage{amsmath,amssymb,amsfonts,amsthm,amscd,bm}
\usepackage{booktabs}
\usepackage{cancel}

\def\tilde{\widetilde}
\def\bar{\overline}

\def\*{\star}
\def\[{\left[}
\def\]{\right]}
\def\({\left(}      

\def\){\right)}

\def\frac#1#2{\dfrac{#1}{#2}}
\def\inv#1{\dfrac{1}{#1}}
\def\half{\tfrac{1}{2}}
\def\d{\partial}

\def\2pi{\hbox{$2\pi i$}}

\def\dsl{\raise.15ex\hbox{/}\kern-.57em\partial}
\def\Dsl{\,\raise.15ex\hbox{/}\mkern-.13.5mu D}
\def\CD{{\cal D}}

      \def\CO{{\cal O}}

\def\2pi{\hbox{$2\pi i$}}

\def\dsl{\raise.15ex\hbox{/}\kern-.57em\partial}
\def\Dsl{\,\raise.15ex\hbox{/}\mkern-.13.5mu D}
\font\numbers=cmss12
\font\upright=cmu10 scaled\magstep1
\def\stroke{\vrule height8pt width0.4pt depth-0.1pt}
\def\topfleck{\vrule height8pt width0.5pt depth-5.9pt}
\def\botfleck{\vrule height2pt width0.5pt depth0.1pt}
\def\Zmath{\vcenter{\hbox{\numbers\rlap{\rlap{Z}\kern
    0.8pt\topfleck}\kern 2.2pt
    \rlap Z\kern 6pt\botfleck\kern 1pt}}}
\def\Qmath{
    \vcenter{\hbox{\upright\rlap{\rlap{Q}\kern3.8pt\stroke}\phantom{Q}}}}
\def\Nmath{\vcenter{\hbox{\upright\rlap{I}\kern 1.7pt N}}}
\def\Cmath{\vcenter{\hbox{\upright\rlap{\rlap{C}\kern
                   3.8pt\stroke}\phantom{C}}}}
\def\Rmath{\vcenter{\hbox{\upright\rlap{I}\kern 1.7pt R}}}
\def\Z{\ifmmode\Zmath\else$\Zmath$\fi}
\def\Q{\ifmmode\Qmath\else$\Qmath$\fi}
\def\N{\ifmmode\Nmath\else$\Nmath$\fi}
\def\C{\ifmmode\Cmath\else$\Cmath$\fi}
\def\R{\ifmmode\Rmath\else$\Rmath$\fi}
\def\barray{\begin{eqnarray}}
\def\earray{\end{eqnarray}}
\def\beq{\begin{equation}}
\def\eeq{\end{equation}}

\def\AA{\leavevmode\setbox0=\hbox{h}
\dimen0=\ht0 \advance\dimen0 by-1ex\rlap{\raise.67\dimen0\hbox{\char'27}}A}

\def\Arg{{\rm Arg}\,}

\makeatletter
\def\iddots{\mathinner{\mkern1mu\raise\p@
\vbox{\kern7\p@\hbox{.}}\mkern2mu
\raise4\p@\hbox{.}\mkern2mu\raise7\p@\hbox{.}\mkern1mu}}
\makeatother

\theoremstyle{plain}

\newtheorem{proposition}{Proposition}

\theoremstyle{remark}

\def\arg{{\rm arg}}
\def\Arg{{\rm Arg}}
\def\half{\tfrac{1}{2}}

\def\rhozero{\rho_\bullet}
\def\sigmazero{\sigma_\bullet}
\def\tzero{t_\bullet}
\def\thetaRS{\vartheta}

\def\charphi{\varphi} 

\def\character{X}

\def\r{r} 

\def\q{q}

\def\ZpZ{\Upsilon}

\def\p{p}  
\def\pprime{\mathfrak{p}}
\def\r{r}                     

\def\ZInteger{\mathbb{Z}}

\def\tstar{t_\star}

\def\rhoWigner{\rho_{\rm w}}

\renewcommand{\thefootnote}{\roman{footnote}} 

\begin{document}

\title{
Spectral Flow for the Riemann zeros }
\author{Andr\'e  LeClair\footnote{Andre.LeClair@cornell.edu}}
\affiliation{Cornell University, Physics Department, Ithaca, NY 14850, USA} 
\renewcommand{\thefootnote}{\alph{footnote}}

\begin{abstract}
Recently,   with Mussardo we defined a quantum mechanical problem of a single particle scattering with impurities wherein 
the quantized energy levels $E_n (\sigma)$  are exactly equal to the zeros of the Riemann $\zeta (s)$ where $\sigma = \Re (s)$ in the limit
$\sigma \to \half$.    The S-matrix is based on the Euler product and is unitary by construction,   thus the underlying hamiltonian is hermitian 
and all eigenvalues must be real.    Motivated by the Hilbert-P\'olya idea we study the spectral flows for $\{ E_n (\sigma) \}$.    This leads to a simple criterion for the validity of the Riemann Hypothesis.      
The spectral flow arguments are simple enough that we present analogous results for the Generalized and Grand Riemann Hypotheses. 
We also illustrate our results for a counter example where the Riemann Hypothesis is violated since there is no underlying unitary S-matrix due to the lack of an Euler product.

\end{abstract}

\maketitle
\tableofcontents

\section{Introduction}

There have been  few approaches to the Riemann Hypothesis (RH) based on ideas from physics.     
Although Riemann's zeta function $\zeta (s)$ is ubiquitous in Statistical Mechanics,  and in Feynman diagram perturbation theory,   in this context this usually only involves $\zeta (s)$ with $s$ equal to an integer,   however one needs a physical problem which involves $s$ in the entire complex plane. 
For a review of some  approaches based on physics up to the year 2011 see for instance \cite{Schumayer}.   
The most prominent physical proposal  of the past is probably the Hilbert-P\'olya idea which turns the problem of establishing the validity of the Riemann Hypothesis into the existence of a single particle quantum  {\it hermitian} Hamiltonian whose bound state eigenvalues are equal to the ordinates of zeros on the critical line  $\Re(s) = \half$ \cite{BerryK,BerryK2,Sierra,Sredincki,Bender}.  The idea is that  a non-hermitian hamiltonian would imply 
complex zeros off the critical line.   A primary motivation for the latter work was the Montgomery-Odlyzko conjecture that the eigenvalue statistics satisfies random matrix GUE statistics for hermitian hamiltonians \cite{Montgomery,Odlyzko}.     
It is fair to say that such a hypothetical hamiltonian has remained elusive,   and no specific model has been proposed from which one can calculate the actual zeros.  

 A different approach, based on statistical physics and random walks, has recently been pursued \cite{LMDirichlet,MLRW,FL2}.   There the validity of 
 the RH is  related to random walks over  the primes.    In this approach one can only prove that the RH is true with probability one,   since it relies on the pseudo-randomness of the primes which is difficult to prove.      One nice feature of this approach is that the same reasoning applies to all the $L$-functions based on non-principal Dirichlet characters,  including  those based on (cusp) modular forms \cite{FL2}.      
 
 This paper is focussed in on the Hilbert-P\'olya idea for the model in \cite{LecMussDefect}.   
In our recent work with Mussardo \cite{LecMussDefect},   we proposed a quantum mechanical scattering problem where the quantized energies on a circle correspond to the Riemann zeros in a certain limit.     The model consists of a single particle moving on a circle with impurities.  
It is defined by a real dispersion relation,  namely the relation between the momentum $p$ and energy $E$ for a single particle,   and the 
S-matrix for the scattering with the impurities.     The model is thus essentially defined by its S-matrix.     
The connection will the Hilbert-P\'olya conjecture is that the S-matrix is $S=e^{-i H}$ where $H$ is a hermitian hamiltonian,  thus
$S^\dagger S = 1$.\footnote{This is an over-simplification, since  in general one has to deal with time-ordered product of the exponential operator.}
 The S-matrix is unitary for this model by construction since it is built from a product of scattering phases based on the Euler product,   where each impurity is associated with a prime number.   Thus the model has an 
implicit  underlying hermitian hamiltonian.
 The main difference with the original version of  the Hilbert-P\'olya idea is that for the latter the quantized energies were imagined  to be bound states rather than scattering states.    
For our  model there is a quantization condition for the energies of the system which comes from a Bethe Ansatz equation,
and one can thereby readily compute the Riemann zeros accurately  from the model,    in contrast to some other proposals based on 
Hilbert-P\'olya.     
The present article continues to study this model,  which we will refer to as the LM model.

Let $\rho_n = \half  + i t_n$ denote a zero of Riemann's zeta function $\zeta (s)$ on the upper critical line $t_n > 0$.    In the LM model,  quantized energies  $E_n (\sigma)$ are functions of  the abscissa  $\sigma$
of the complex variable $s = \sigma + i t$.   These $E_n (\sigma)$ are eigenvalues of a quantum mechanical problem  and don't correspond to zeros of a complex analytic function,  except in the limit: 
\beq
\label{tnEn}
t_n = \lim_{\sigma \to \half^+}  E_n (\sigma).
\eeq
It is thus natural to reconsider the Hilbert-P\'olya idea for the LM model,    and this is the main focus of this article.    
The advantage of the LM model is the dependence of the quantized energies on $\sigma$.    
    One is thus  led to consider the spectral flow,  i.e. how $E_n (\sigma)$ depends on $\sigma$.     For the LM model, all eigenvalues
    $E_n (\sigma)$ should be real  since the underlying hamiltonian is hermitian,   and as we will see this will lead a simple criterion for the 
    Riemann Hypothesis (RH) to be true.

  \begin{figure}[b]
\centering\includegraphics[width=.5\textwidth]{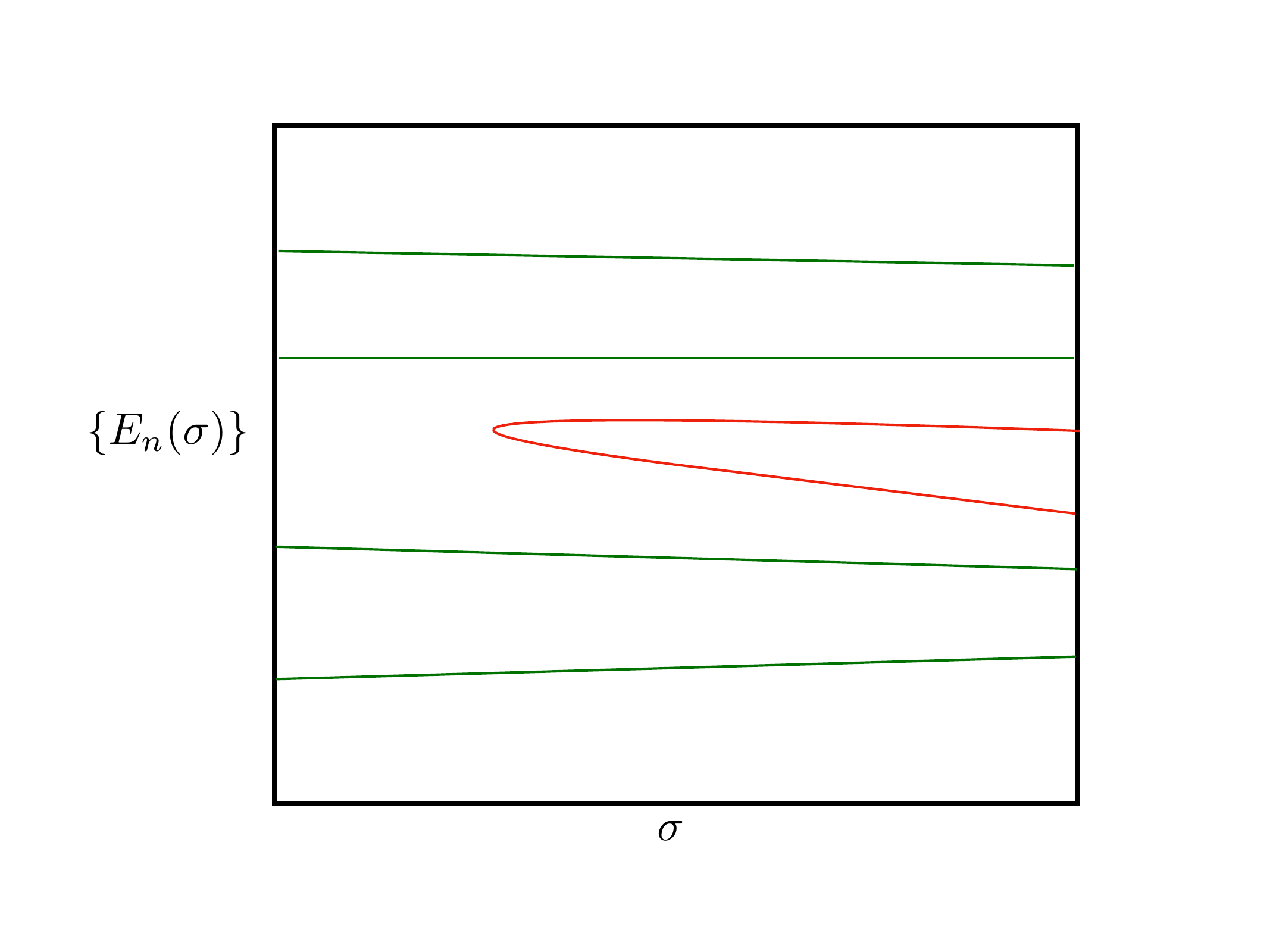}
\caption{A generic situation for eigenvalues of non-hermitian matrices:    two real eigenvalues  coalesce to form a pair of complex conjugates.}
 \label{Offline}
\end{figure}

     Relatively recently there has been a lot of interest in non-hermitian quantum mechanics in connection with applications to open quantum systems
 and their Lindbladian dynamics,   and some of the ideas of the present work were influenced by this literature.       Random matrices for non-hermitian hamiltonians were classified in \cite{BL38}   of which there are 38  universality classes.\footnote{These are now commonly referred to as the Bernard-LeClair classes.}       
 Generically, if the $\{ E_n (\sigma) \}$ are related to a {\it non-hermitian} hamiltonian,   then a pair of real eigenvalues can coalesce to become a pair of non-real eigenvalues which are complex conjugates.     See Figure \ref{Offline}.    
  Although the literature by now is quite large,   let us just mention a few  works that are perhaps more relevant  to this paper \cite{Ryu,Ribiero,Verbaarschot}.    In the latter \cite{Verbaarschot},   at least $14$ of the $38$ classes were realized,
 and  one can find plots for the SYK model \cite{SYK1,SYK2} exhibiting the generic features displayed in Figure \ref{Offline}.  
 These works are just a motivation and the present work does not rely on this work in detail.\footnote{We should mention that there are some 
 superficial similarities between the SYK and LM models,  in that the interactions are random.    In the LM model the randomness stems from the
 prime numbers that define the S-matrix.}    
 However below we will study a Dirichlet $L$-function which  corresponds to an underlying non-hermitian hamiltonian and where the RH is thereby false.

We present our work as follows.    In the next section we review  the  LM model  and it's Bethe-Ansatz equations for $\{ E_n (\sigma) \}$.    
In Section III we derive the spectral flow equations.     This leads to a simple conditions for there to be no complex zeros off the line,  which we refer to as Propositions.    We present compelling numerical evidence for these propositions.      
As a test of these ideas it is important to have a counter example which shows how the RH can fail,  and this is presented in Section IV. 
In Section V we show how our propositions are consistent  with known results in analytic number theory if one assumes the RH.    
Since the spectral flow argument so readily leads to equivalences to the RH,   in Section VI  we  present extensions of our propositions to the two infinite classes of 
$L$-functions:   those based on Dirichlet arithmetic characters and those based on (cusp) modular forms.    In Section VII we study random matrix statistics for eigenvalues $E_n (\sigma)$ off the critical line.   
Results from analytic number theory we need can all be found in the  elementary books \cite{DavenportBook,Apostol}. 
For mathematical physicists, our  review lectures \cite{Hanover} may also be useful since they are condensed to include all that is needed here and  are perhaps more approachable.

\section{The model and it's Bethe Ansatz equation}

\def\Egap{E_{\rm gap}}

In this section we review the LM model,  omitting some details  that can be found in \cite{LecMussDefect}. 
Consider a single particle of momentum $\p$ moving on a circle of circumference $R$.  
We need to specify a dispersion relation $\p (E)$ where $E$ is the energy of a single particle. For now,    we chose this dispersion relation to be
\beq
\label{Dispersion2}
\p (E)  = E \, \log \( \frac{E}{2 \pi  e } \). 
\eeq
 In  the infinite interval for $E > \Egap \equiv 2 \pi e \approx 17.08$,   the above dispersion relation can be inverted as 
\beq
\label{Eofp}
E(\p)  = \frac{p}{ W(\p/2 \pi  e )}
\eeq
where $W$ is the principal branch of the Lambert $W$ function.    Since $E(\p)$ must be greater than $2 \pi e$,   this dispersion relation 
has an intrinsic energy gap $\Egap$.      
The gap is obvious from the small $p$ expansion,
\beq
\label{EofpSmall}
\frac{E}{2 \pi e} = 1 + \frac{p}{2 \pi e} - \frac{p^2}{8 \pi^2 e^2} + \frac{p^3}{12 \pi^3 e^3} + \CO (p^4).
\eeq
This shows that above the gap $E(p)$ can be interpreted as corrections to a relativistic dispersion.    
For large $p$, 
$E(\p) \approx  \p/ \log \p$,  which would be relativistic if it weren't for the $1/\log p$.   
This is not a standard dispersion relation,   however many uncommon dispersion relations arise in condensed matter physics. 
 For our purposes,   \eqref{EofpSmall}  implies that the hamiltonian is hermitian since $p$ is a hermitian operator:   in position space it is
 the usual $p = -i \hbar \d_x $.

\begin{figure}[t]
\centering\includegraphics[width=.7\textwidth]{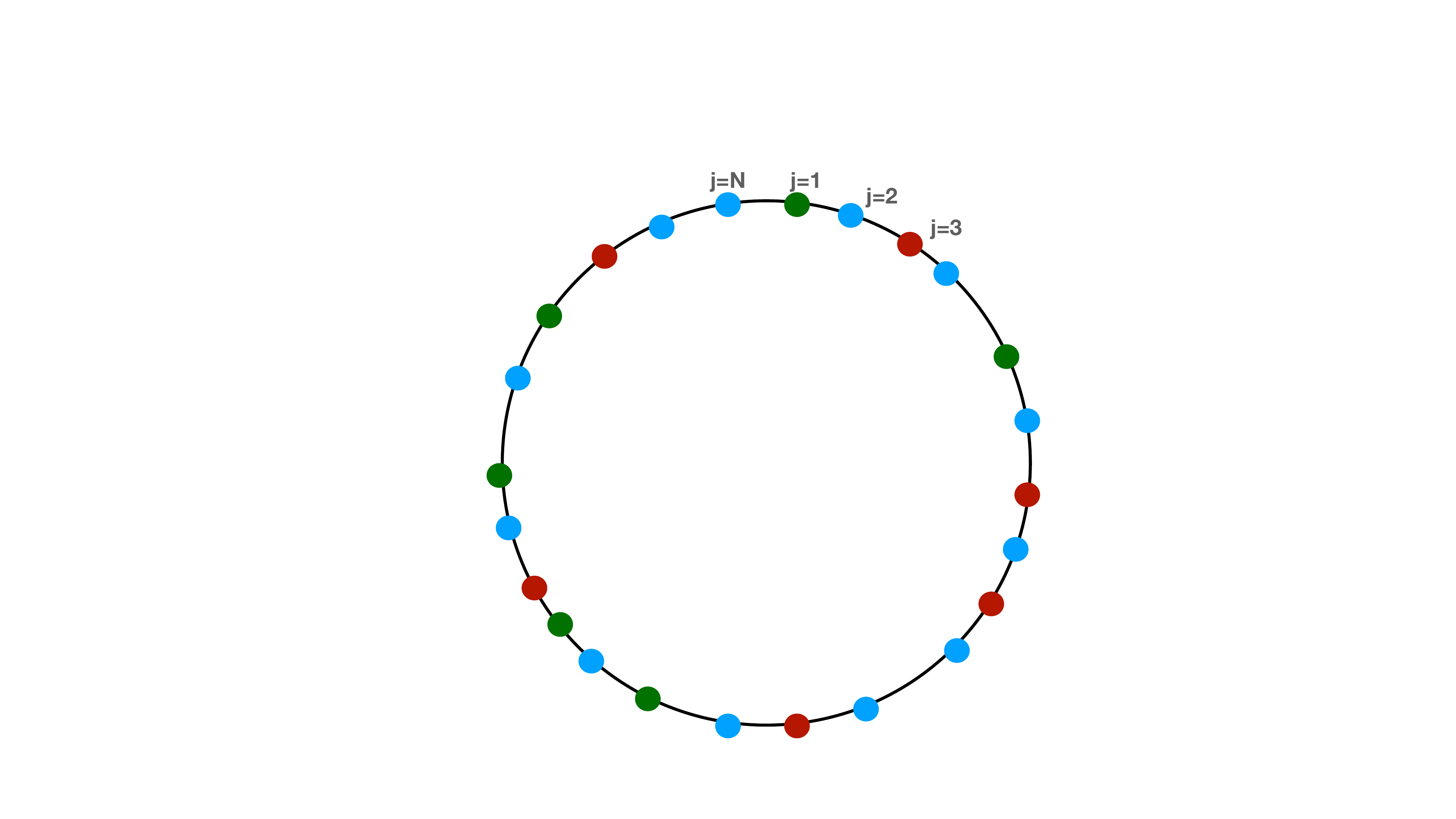}
\caption{Impurities  on a circle labeled $j=1, 2,..., N$.   Different colors denote different scattering phases.}
 \label{DefectFigure}
\end{figure}

Consider  $N$ stationary impurities spread out on the circle,  with no particular location,  except that they are separated,  and label them $j = 1, 2, \ldots N$,  as illustrated in Figure \ref{DefectFigure}.    We assume there is no reflection,   namely the scattering is purely transmitting. There are many known examples of  purely transmitting relativistic theories \cite{DMS,KonikLeClair,Corrigan},  in fact infinitely many that are integrable, and there are also non-relativistic examples of reflectionless potentials \cite{NonRelativistic0,NonRelativistic}.  
To each impurity labeled $j$ we associate a transmission S-matrix $S_j (\p)$,  which by unitarity, is a phase
\beq
\label{Si}
S_j (\p ) = e^{i \phi_j (\p)} \,\,\,.
\eeq
Due to the purely transmitting property,  the  scattering  matrix for 2 impurities $j,j'$ is simply  $S_j (\p) S_{j'}(\p) $,  and so forth.     

As the particle moves around the circle, it scatters through each impurity and, coming back to its original position, the matching requirement for its wavefunction leads to the quantization condition of its momentum $p$ expressed by
\beq
\label{Bethe1} 
e^{i \p R} \, \prod_{j=1}^N  S_j (\p) = \pm 1 \,\,\,,
\eeq
where $+1, -1$ corresponds to bosons, fermions respectively (see, for instance \cite{MussardoBook} and references therein).   
For our purposes it is necessary to take  the particles to be fermions,  and we end up with the  Bethe-ansatz equation:
\beq
\label{Bethe2}
\p_n  R +  \sum_{j=1}^N   \phi_j (\p_n  ) = 2 \pi  \, (n - \half) \,,
\eeq
for some integer $n$.  When there is no impurity scattering,  namely $S_j =1$ and $\phi_j = 0$,  the above quantization condition is the usual one for particles on a circle that can be found in any elementary Quantum Mechanics book.         Then the quantized energies of the system are $E_n = E(p_n)$, where $p_n$ is the solution of the Bethe-ansatz equation relative to the integer $n$.     The fact that we need to treat the particles as fermions is perhaps  related to the known level-repulsion of energy levels
$E_n$.

\def\r{\pprime} 
\def\p{\pprime}

\def\q{\pprime}

Thus far the model is quite general.      For our purposes we define the S-matrices as follows.  
   We assume that the transmission S-matrices are more easily expressed in terms of the energy $E$ rather than  the momentum $p$.   
To each impurity $j=1,2, \ldots$  we associate  the $j$-th prime number $\pprime_j$,~   
$\{\pprime_1,  \pprime_2, \ldots\} = \{2,3,5, \ldots\}$,   and a constant  real phase angle $\charphi_j$:
\beq
\label{Sjs}
S_j (E) = \frac{\q_j^{\sigma}    - e^{i (E \log \q_j - \charphi_j )}  }
{ \q_j^{\sigma}   - e^{- i (E \log \q_j  - \charphi_j )}   } ,
\eeq
where $\sigma$ is a free parameter which is a  positive real number which we will eventually limit to $\sigma > \half$.       
The phases $\charphi_j$ are relevant for more general $L$-functions based on Dirichlet characters.  See below for an example  based on a 
mod 5 character.      
For the Riemann zeta case,  we chose $\charphi_j =0$.
 This implies that the scattering phases are 
\beq
\label{phis}
\phi_j (E) =  - 2 \, \Im \log \(  1 -  \frac{e^{- i E \log \q_j  } } { \q_j^{\sigma} }  \)\, .
\eeq
Without loss of generality, we set $R=1$,  since it can be recovered by scaling the momenta $p$.  The Bethe equations \eqref{Bethe2} read 
\beq
\label{Bethe3}
\frac{E_n }{2} \log \( \frac{E_n }{2 \pi  e} \)  -  \sum_{j=1}^N   \Im \log \(  1 -  \frac{e^{- i E_n \log \q_j  } } { \pprime_j^{\sigma} }  \) = (n - \tfrac{3}{2} ) \pi
\eeq
where we have shifted for later convenience $n$ by $-1$.\footnote{This can be traced to the existence of a pole in $\zeta (s)$ at $s=1$.} 
The  solutions $E_n$'s  obviously depend on $\sigma$, which we denote as $E_n (\sigma)$ unless otherwise indicated.  
Above,  the prime numbers $\pprime$ could be replaced a real number greater than one,    which could be random,   and this still defines an interesting  quantum  scattering problem.

Let us now explain how the eigenvalues $E_n (\sigma)$ are related to the Riemann zeros.  
Let us introduce the complex variable $s= \sigma + i E$.   
 If $\sigma>1$, 
 we can use the Euler product for the Riemann zeta function 
 \beq
\zeta(s) \,=\,\sum_{n=1}^{\infty} \frac{1}{n^s} \,=\,\prod_{j=1}^{\infty} \(1 - \frac{1}{\p_j^s}\)^{-1}  ~~~~~(\Re (s) > 1 ).
\label{zetaRiemann}
\eeq
to sum over scattering phases in the thermodynamic limit $N \to \infty$.     
  The equation becomes 
\beq
\label{FLasymptotic} 
\frac{E_n }{2} \log \(\frac{ E_n }{2 \pi  e }  \)  + \arg \, \zeta (\sigma + i E_n )= (n - \tfrac{3}{2} ) \pi .
\eeq
For $\sigma > 1$ the thermodynamic limit is well defined since the Euler product converges.   
As is common in quantum mechanical problems,   for $\sigma < 1$ we define the sum over scattering phases via analytic continuation,  
and this thermodynamic limit through analytic continuation will be henceforth assumed.   
It is important that $\arg$ in the above equation is not the principle branch $\Arg$;   $\arg$ is a sum of the $\Arg$'s in \eqref{phis},  but they 
can accumulate.        In fact it is known that the above $\arg$ is unbounded,  being roughly $\log E$.   
If one neglects the $\arg \, \zeta$ term,  then 
for large $n$ the solution to the equation (\ref{FLasymptotic}) above is approximately $E_n \approx \tilde{E}_n$ where 
\beq
\label{EnLambert}
\tilde{E}_n  \approx \frac{2 \pi (n-\tfrac{11}{8})}{W\( (n- \tfrac{11}{8})/e \)}
\eeq
where $W$ is the principle branch of the Lambert function.   The $\tilde{E}_n$ should be viewed as the quantized energies of the free theory 
with the S-matrix equal to 1.

The above equation \eqref{FLasymptotic} was first proposed in \cite{Electrostatic},   and    
 is an excellent approximation to the Riemann zeros on the critical line as $\sigma \to \half^+$.   
 See \eqref{En100} below.    
The reason for this is quite simple as we now explain.  
 Following standard conventions in analytic number theory,  let  us now define a 
 complex variable $s= \sigma + i t$ where, based on the notation above, 
  $t= E$.  Henceforth we will use $E,t$  interchangeably if there is no cause for confusion. 
   We consider zeros {\it on the critical line},  which are known to be infinite in number. Denote the $n$-th zero 
 on the upper critical line as
\beq
\label{rhozero}
\rho_n = \half + i t_n, ~~~~~n=1,2,3, \ldots,    ~~~\zeta (\rho_n ) =0
\eeq
where $t_1 = 14.1347..$ is the first zero, and so forth. 
Define a completed $\zeta$ function which satisfies the functional equation as follows:
\beq
\label{chidef}
\chi (s) = \pi^{-s/2} \Gamma(s/2) \zeta (s),  ~~~~~
\chi(1-s) = \chi(s).
\eeq
Furthermore,  define its argument $\theta$:
\beq
\label{thetasig}
\theta (\sigma,t) = \arg\, \chi (\sigma + it), ~~~~~~
\theta (\sigma, t)  = \thetaRS(\sigma,t) + \arg \, \zeta (\sigma+ i t) ,  ~~~~~~~~
\thetaRS(\sigma, t) \equiv \arg \, \Gamma \( \half (\sigma + i t) \) - \tfrac{t}{2} \log \pi.
\eeq
On the critical line $\thetaRS (\half, t)$ is smooth and  commonly referred to as the Riemann-Siegel $\thetaRS$ function.  
Below,  if it is implicit that we are on the critical line we will simply write $\thetaRS (t) \equiv \thetaRS (\half, t)$.  
As argued in \cite{LecMussDefect},    since $\chi(s)$ is real on the critical line,  it must jump by $\pi$ at each simple zero.   
This implies that the exact Riemann zeros on the line
\beq
\label{tnEn}
t_n =  \lim_{\sigma \to \half^+} \,  E_n (\sigma),
\eeq
satisfy the equation \cite{FrancaLeClair,LecMussDefect}:
\beq
\label{Horzn}
\lim_{\delta \to 0^+ } \theta (\half + \delta, t_n ) = (n-\tfrac{3}{2})\pi,   ~~~~\Longrightarrow ~~~
\thetaRS (t_n) +  \lim_{\delta \to 0^+}  \arg \, \zeta(\half + \delta + i t_n )  = (n - \tfrac{3}{2} ) \pi .
\eeq

Let us now return to the equation \eqref{FLasymptotic}.   One easily sees that the $\thetaRS (t_n)$ term in \eqref{Horzn}  amounts to small corrections to our dispersion relation \eqref{Dispersion2}:
 \beq
 \label{DispersionRS}
   p(E)  \approx  2 \thetaRS(E) = E \, \log \( \tfrac{E}{2 \pi e}\) - \frac\pi4 + \frac{1}{24 E} + O(1/E^3 ).  
  \eeq
As for $p(E)$ in eq.\,(\ref{Dispersion2}), $\thetaRS (E)$ is monotonic and positive in the infinite interval $(E_*,\infty)$,  where 
$E_* \simeq 17.8456$ is a minor deformation of the gap $\Egap = 2 \pi e$,  and in this interval it is invertible and therefore asymptotically eq.\eqref{Eofp} is valid.  
 Thus 
  \eqref{FLasymptotic} should be viewed as a large $t$ asymptotic version of 
  the exact equation \eqref{Horzn}.       However for reasons already mentioned,   it is quite accurate for relatively low $t$,   and 
  \eqref{FLasymptotic}  provides excellent results for even the first few zeros.     
 For instance from  \eqref{FLasymptotic}  with $\delta = 10^{-6}$ we obtain
\beq
\label{En100} 
\lim_{\sigma \to \half^+}  \{ E_{1} (\sigma), E_{2}(\sigma) , \ldots , E_{5}(\sigma) \}  =
\{14.134724, ~21.022039,~ 25.010857, ~
30.424875, ~32.935061 \},
\eeq
which are identical  to the true Riemann zeros to 8 digits.   We emphasize that the exact equation \eqref{Horzn} yields 
zeros to arbitrary precision by systematically reducing $\delta$,  as illustrated in \cite{FrancaLeClair}.

   Henceforth,  we will work mainly  with the asymptotic equation
  \eqref{FLasymptotic} since the $1/E$ corrections in \eqref{DispersionRS} are either negligible for high enough $E$ or easily incorporated without affecting the main results.  Such corrections  just make the formulas more complicated in an irrelevant and uninteresting  manner for our purposes since they only matter at very low $t$ where the RH is already known to be valid.\footnote{See for instance equation \eqref{Dave2} where we display such corrections in that case.}     Thus it is implicit in the Propositions below that $t> \tstar$ where for $t>\tstar$ one can neglect the  $\CO (1/t)$ corrections coming from 
  $\thetaRS (t)$.   It is not difficult to determine $\tstar$ on a case by case basis.   
  One can roughly estimate $\tstar$ as the value of $t$  above which $\thetaRS (t) >0$, 
  $\thetaRS (\tstar) \approx  0$.
      Here $\tstar = 17.8456..$,  which is just above the first zero,  and is actually an overestimate,  as can be seen in Figure
      \ref{PropHalfLowT}.      In any case,  for all $L$-functions considered in this paper, 
     $\tstar$ is low and typically less than $20$.   
   Thus, below  the symbols $\lesssim$ and $\simeq$  refer merely to neglecting these $\CO(1/E)$ corrections, namely it is assumed that 
    $t> \tstar$ where $\tstar$ is quite low and in a region where zeros can be easily ruled out numerically.       
 The justification for this will be clear from the numerical results below.

  \bigskip
  
 The above considerations lead to our first Proposition:

\begin{proposition}
\label{Prop0}
Since the S-matrix for the LM model,    which is based on the Euler product formula,  is unitary,   and the non-interacting part of the hamiltonian
is hermitian based on \eqref{EofpSmall},  
all eigen-energies $E_n (\sigma)$ are real in the thermodynamic limit with $\sigma > \half$.   
\end{proposition}

\bigskip

The LM model is mainly defined by it's unitary S-matrix based on the Euler product.     Although we didn't present a formula for the implicit underlying hamiltonian,   there are many examples of such theories defined by their S-matrix,     namely where the theory is defined by its S-matrix and an explicit hamiltonian is unknown. 
Henceforth we assume  the theory is well-defined by its dispersion relation and it's unitary S-matrix.   

  In \cite{LecMussDefect}  we argued that the RH is true if $\chi (s)$ is a regular alternating function,  however after some attempts this seems difficult to prove.       The present article will present much more succinct criteria for validity of  the RH.

  \bigskip
  
 Let us mention a  relevant remark since it is one of the primary motivations for this article.         One has the following theorem \cite{FrancaLeClair}: 
   
   \bigskip

 \noindent $\bullet$ ~  {\it  If for every integer $n>0$ there is a unique solution of the equation \eqref{Horzn}  for $E_n (\sigma)$  in
 the limit $\sigma \to \half^+$,  then the Riemann Hypothesis is true and all zeros are simple. }   

\bigskip

The reason this statement  is correct is quite simple:    If there is a solution for every $n$,  then the zeros along the critical line are enumerated and their counting saturates the known formula for the total number of zeros in the critical strip upto  height $t<T$,  commonly
 referred to as $N(T) = \thetaRS (T)/\pi   + \arg \, \zeta(\half + i t)/\pi  + 1 $.   
  Clearly any zeros off the critical line will spoil this.    In Section IV we show how this is violated for a counterexample.   
 
\bigskip

\section{Spectral flow and a simple criterion for the validity of the Riemann Hypothesis}

\def\sigmastar{\sigma_*}

\subsection{The spectral flow equation}

As in the last section,   we will use $E$ and $t$ interchangeably. 
We will  mainly work with the asymptotic form \eqref{FLasymptotic} rather than the exact \eqref{Horzn} since order $1/E_n$ corrections are 
easily neglected as explained above.   This will be implicit in some formulas below,   where by ``large" $t$  we mean $t\gtrsim 20$.

\begin{figure}[t]
\centering\includegraphics[width=.7\textwidth]{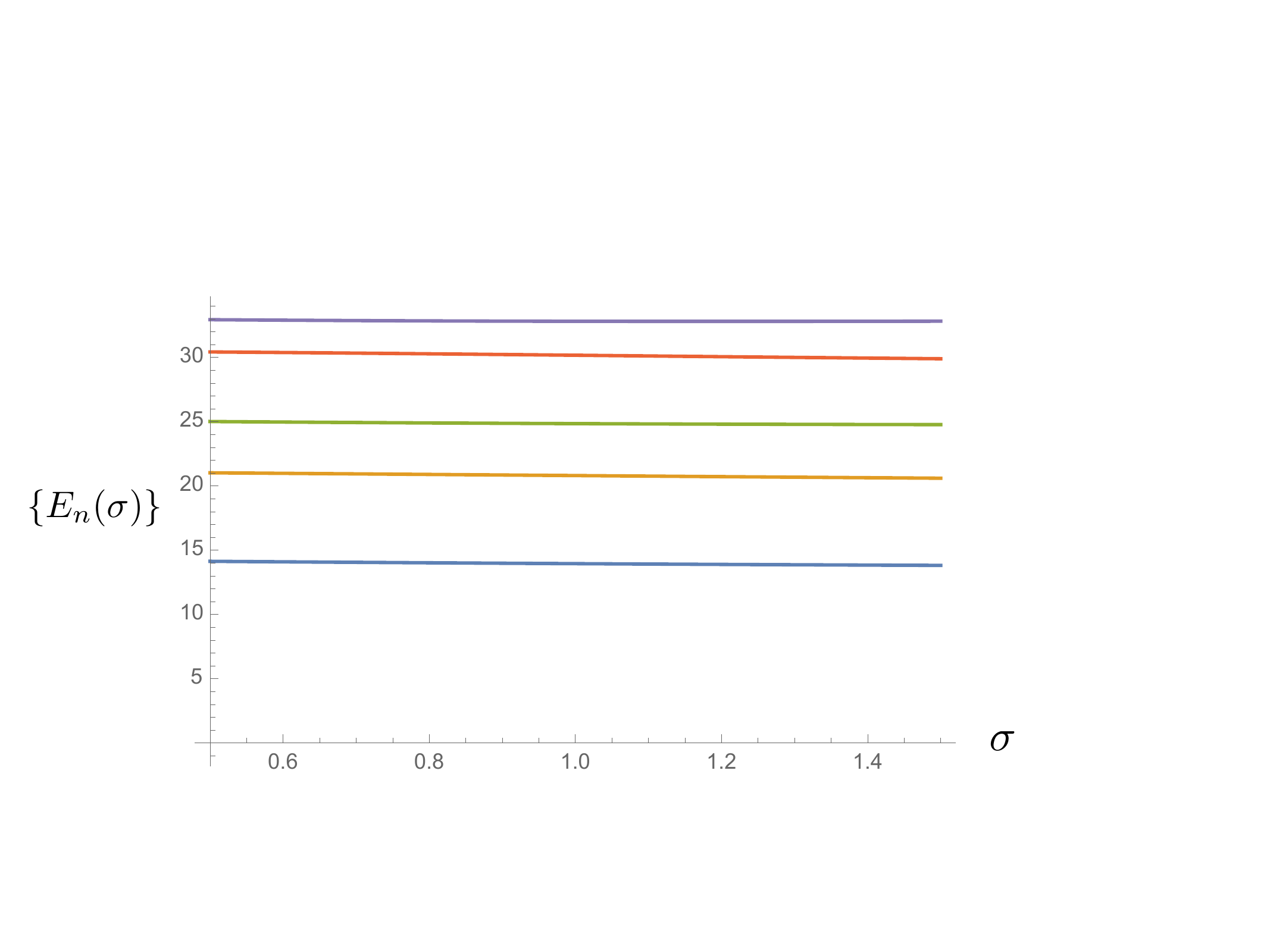}
\caption{A plot of $E_n (\sigma)$ for $n = 1,  2, \ldots, 5$ as a function of $\sigma$  to the right of the critical line.}
 \label{EnOfSig5}
\end{figure}

For $\sigma > 1$ there is a unique solution  $E_n (\sigma)$ to  equations \eqref{Horzn} and \eqref{FLasymptotic} for each $n$ since the $\arg \, \zeta$ term is well defined 
through the Euler product and the left hand side is monotonic.     In this section we are concerned with the dependence of $E_n (\sigma)$ on $\sigma$ 
as one moves into the critical strip $\half  < \sigma < 1$ starting from $\sigma>1$ where the $E_n (\sigma)$ are well-defined.    
In Figure \ref{EnOfSig5}  we show this dependence on $\sigma$ for the lowest five zeros.  
One sees that the $E_n$ smoothly deform to the actual zeros $t_n$ as $\sigma \to \half$ without a great deal of variation nor drama.  
Compare with Figure \ref{LevelCrossing} below,  where as we will see the RH is violated in a well-known counterexample.        This implies the zeros $t_n$  can be roughly 
{\it  estimated}  by $E_n (\sigma)$ with $\sigma>1$ where the Euler product converges.

As motivated in the Introduction,   if there were a zero off the critical line,   one expects a pair of real eigenvalues  to coalesce into a pair of complex 
eigenvalues.       Inspection of Figure \ref{Offline} indicates that $dE_n/d\sigma$  diverges at this point.   
Differentiating \eqref{Horzn} with respect to $\sigma$ one finds
\beq
\label{dEn}
\frac{d E_n (\sigma )}{d\sigma} = - \frac{\Im (\ZpZ (s))}{\Re (\ZpZ (s)) + \thetaRS'(E_n (\sigma) ) }, ~~~~~ s = \sigma + i E_n (\sigma)
\eeq
where $\thetaRS'(E) = \d_E \thetaRS (E)$ and we have defined 
\beq
\label{ZpZDef} 
\ZpZ (s) = \frac{\zeta' (s)}{\zeta(s)}, ~~~~~~ \zeta' (s) = \d_s \zeta (s) .
\eeq
We thus propose that at a zero off the critical line,   the denominator in \eqref{dEn} vanishes.  
This leads us to the proposition:

\bigskip
\begin{proposition}
\label{Prop1}
If for a given $\sigmastar$ and a region $t_1 < t < t_2$,  the following condition is satisfied
\beq
\label{Proposition1}
- \Re ( \ZpZ (s) ) <  \thetaRS' (t), ~~~~~ s = \sigmastar + i t,
\eeq
 then all  $E_n (\sigma)$  with $\sigma > \sigmastar$ in this region of $t$ are real.  
For large  enough $t > \tstar \simeq  20$,  the above is equivalent to 
\beq
\label{conjecture2}
- \Re ( \ZpZ (s) ) \lesssim  \, \inv{2} \log \( \frac{t}{2 \pi} \).
\eeq
Clearly this implies that the RH is true if the condition \eqref{conjecture2} is satisfied for all  $\sigma> \half$ and for all $t>0$.   
\end{proposition}
\bigskip

One can also learn something  by considering the integer $n$ as a continuous variable.   On the critical line,  if the $E_n (\sigma)$ are real solutions,
then they are discrete and $dE_n (\sigma)/dn$  should be $\infty$ between zeros.        Differentiating \eqref{FLasymptotic},  one finds 
\beq
\label{EnDn}
\frac{\d E_n (\sigma)}{\d n} = \frac{\pi}{\Re (\ZpZ (s) ) + \half \log \( E_n (\sigma) /2 \pi  \) },   ~~~~~~ s = \sigma + i E_n (\sigma).
\eeq
The same denominator as in \eqref{dEn} appears.    
Now if the RH is true,  all $E_n (\sigma = \half)$ are real and well-defined  and  $dE_n/dn$ must be $\infty$ between zeros.
     This leads to another proposition:

\bigskip

\begin{proposition}
\label{Prop2}
For any $t$  not equal to a zero $t_n$ on the critical line,   one has the equality:\footnote{
We repeat that it   is implicit that this is for $t$ large enough,  namely $\simeq$ in the above equation signifies we are neglecting 
$\CO(1/t^2)$ corrections which vanish at large $t$ and don't affect our Propositions for $t \gtrsim 20$.}        
\beq
\label{RHequality}
- \Re (\ZpZ (s) )  \simeq  \half \log (t/2\pi), ~~~~~~~~~~s = \half + i t.
\eeq
Since this is an equality and not a bound,  the $1/2\pi$ inside the $\log$ is essential.  
\end{proposition}

\bigskip

Thus far we have not assumed the RH,  and this is of key importance.       In fact,   Propositions  2 and 3 follow from Proposition \ref{Prop0}. 
As we will explain in Section V,    Proposition 2 has already been shown  to be equivalent to the RH  by  Lagarias \cite{Lagarias} using some rather sophisticated  analysis compared to our simple spectral flow argument,  and of course by assuming the RH.\footnote{We discovered this subsequent to the above 
spectral flow argument upon 
searching through the equivalents of the RH listed in \cite{Borwein}.}      In Section V,  we will provide a relatively simple proof of the above Proposition 2 by assuming the RH.

\subsection{Numerical evidence}

In this section we provide some convincing numerical evidence for the above Propositions,  in particular in the forms \eqref{conjecture2}
and \eqref{RHequality}.   

\begin{figure}[t]
\centering\includegraphics[width=.7\textwidth]{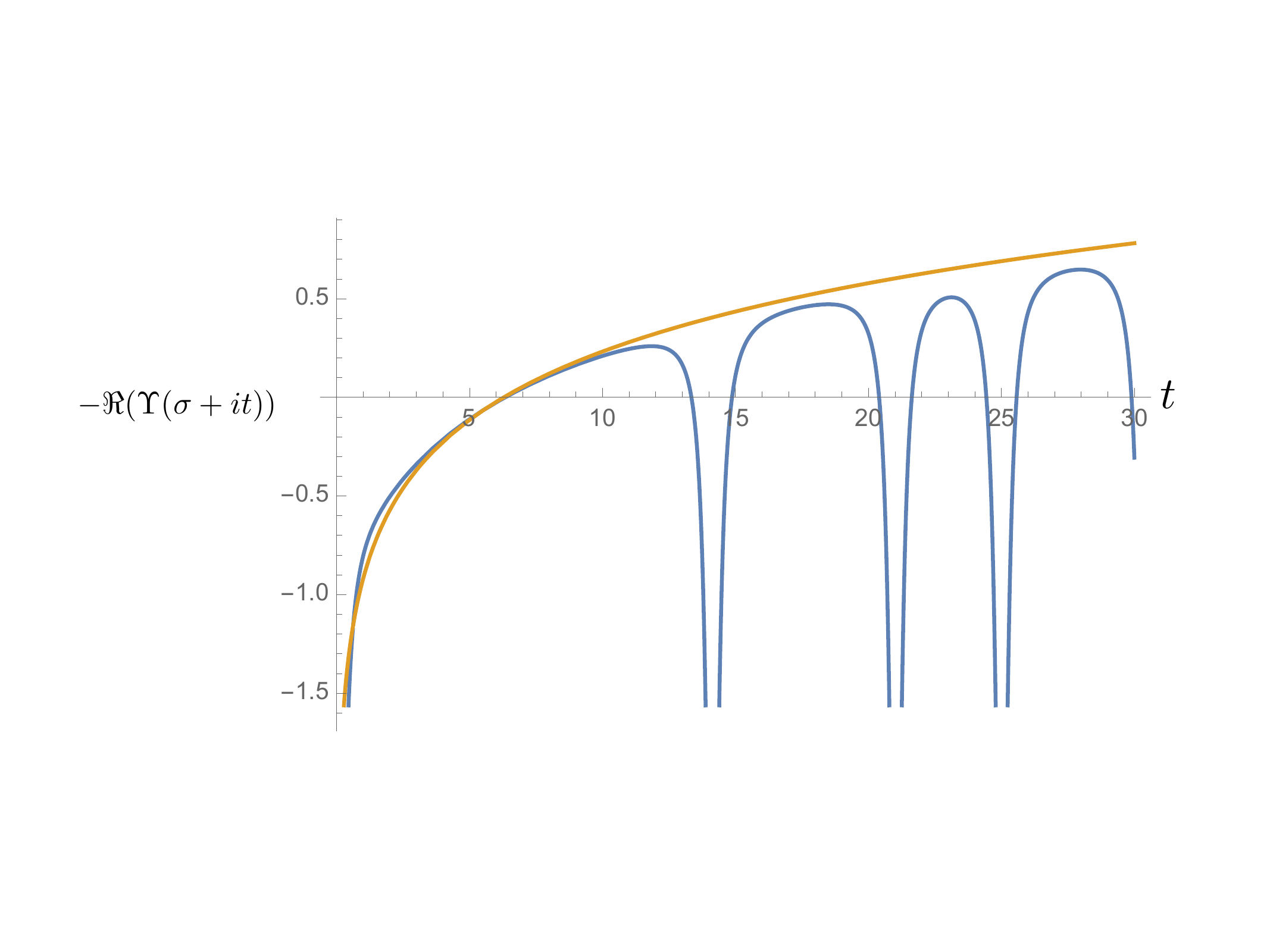}
\caption{A plot $-\Re (\ZpZ (s))$ for $s=\tfrac{3}{4} + i t$ and $\half \log(t/2 \pi)$  as a function of  $t$ for $0< t < 30$.
The asymptotic version of our bound \eqref{conjecture2} is only violated for very low $t \lesssim 10$ which is uninteresting.}
 \label{Prop34LowT}
\end{figure}

\begin{figure}[t]
\centering\includegraphics[width=.6\textwidth]{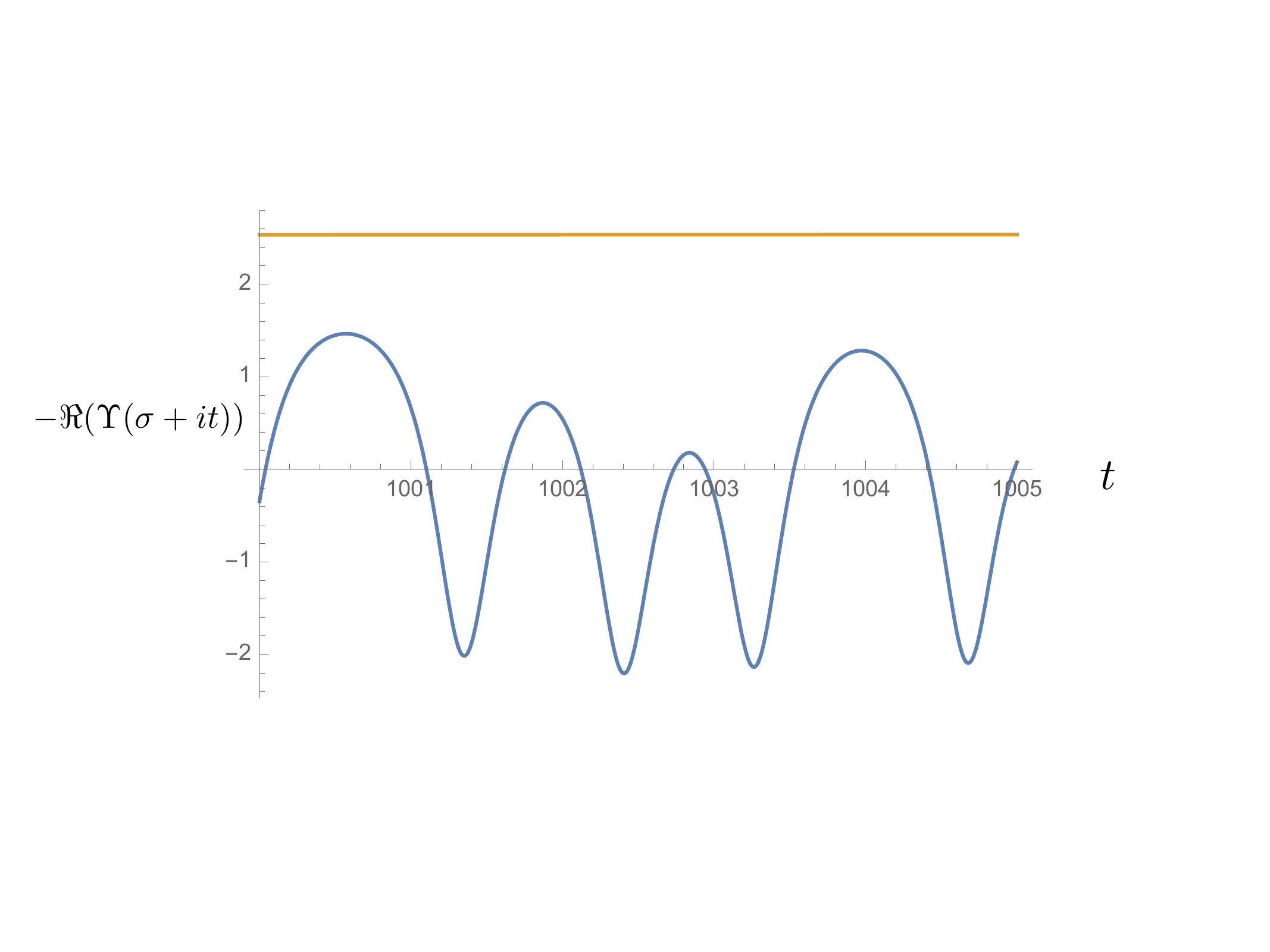}
\caption{A plot $-\Re (\ZpZ (s))$ for $s=\tfrac{3}{4} + i t$ and $\half \log(t/2 \pi)$  as a function of  $t$ for $1000< t < 1005$.}
 \label{Prop34HighT}
\end{figure}

In Figure \ref{Prop34LowT}  we consider $\sigma = \tfrac34$ for very low $t$.    Clearly above the relatively low value $t \sim 10$, the inequality 
\eqref{conjecture2} is satisfied.   It is interesting to consider the same $\sigma = \tfrac34$ for a considerably higher $t \sim 1000$ as shown in 
Figure \ref{Prop34HighT}.      Both these plots rule out zeros for $\sigma > \tfrac34$ in these ranges of $t$,    but interestingly it is easier to rule out such zeros for {\it higher} $t$.

\begin{figure}[t]
\centering\includegraphics[width=.7\textwidth]{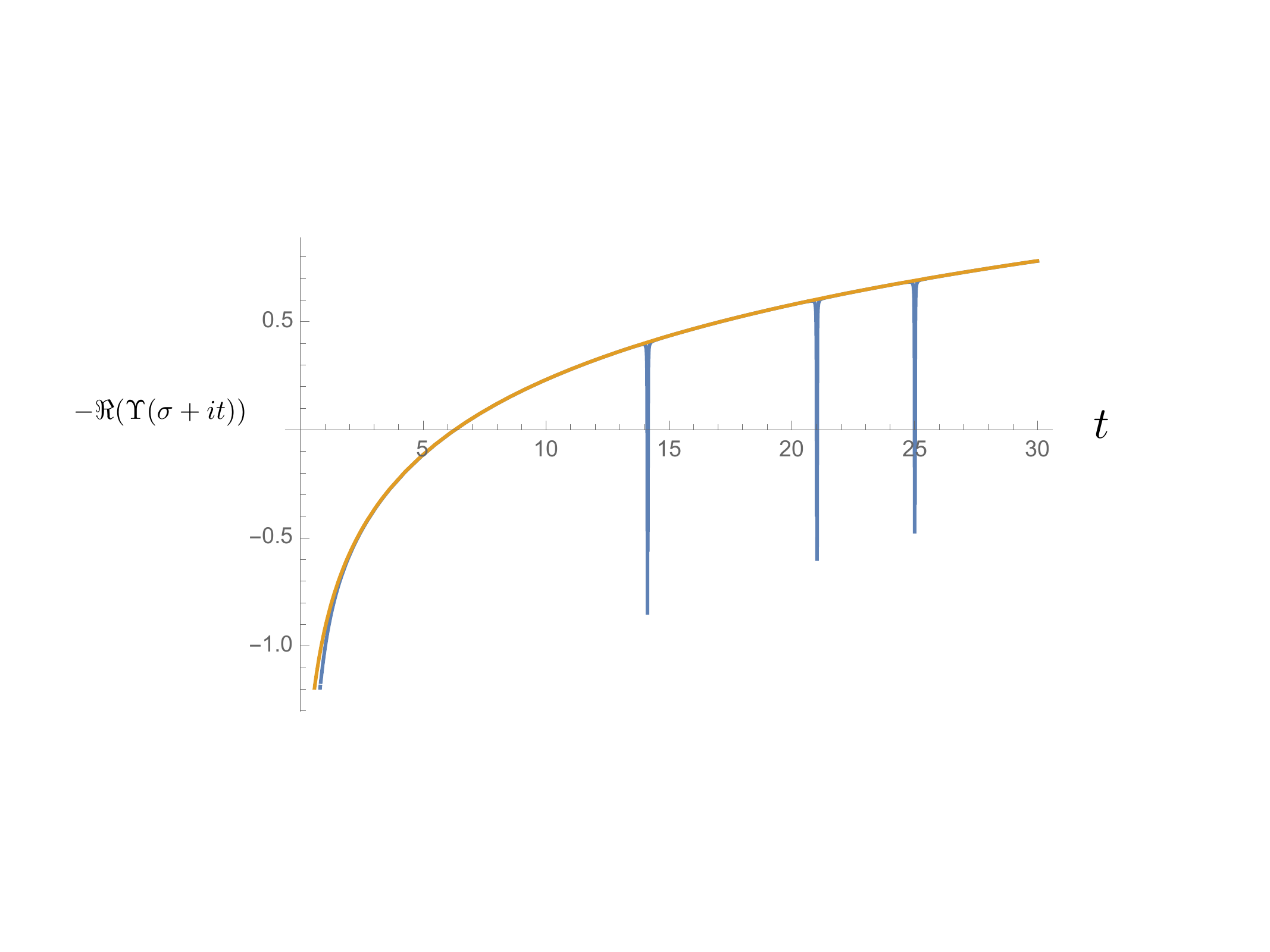}
\caption{A plot $-\Re (\ZpZ (s))$ for $s=\half + \delta  + i t$ with $\delta = 10^{-4}$  and $\half \log(t/2 \pi)$  as a function of  $t$ for $0< t < 30$.}
 \label{PropHalfLowT}
\end{figure}

In Figure \ref{PropHalfLowT}  we approach the critical line with $\sigma = \half + 10^{-4}$.     As expected one gets much closer to the bound
\eqref{conjecture2},  but it is still satisfied.     This pattern repeats itself at larger $t \sim 1000$,   as shown in Figure \ref{PropHalfLowT}.   
What is interesting is that exactly on the critical line,    the zeros are hardly visible and fit equation  \eqref{conjecture2} almost perfectly,  supporting
Proposition \ref{Prop2}.  

 \begin{figure}[t]
\centering\includegraphics[width=.7\textwidth]{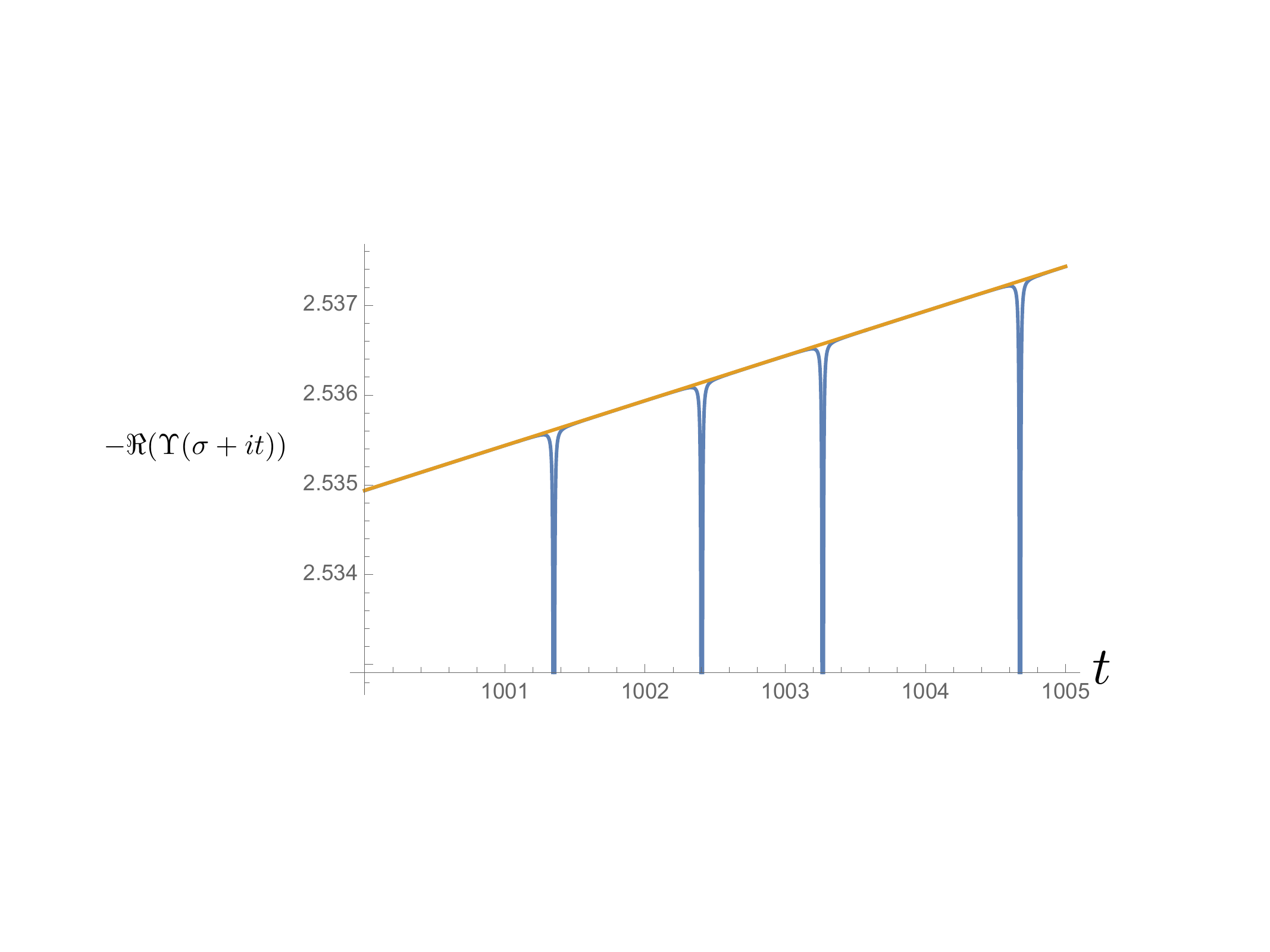}
\caption{A plot $-\Re (\ZpZ (s))$ for $s=\half + \delta  + i t$ with $\delta = 10^{-7}$  and $\half \log(t/2 \pi)$  as a function of  $t$ for $1000< t < 1005$.}
 \label{Proposition1000}
\end{figure} 

\begin{figure}[t]
\centering\includegraphics[width=.6\textwidth]{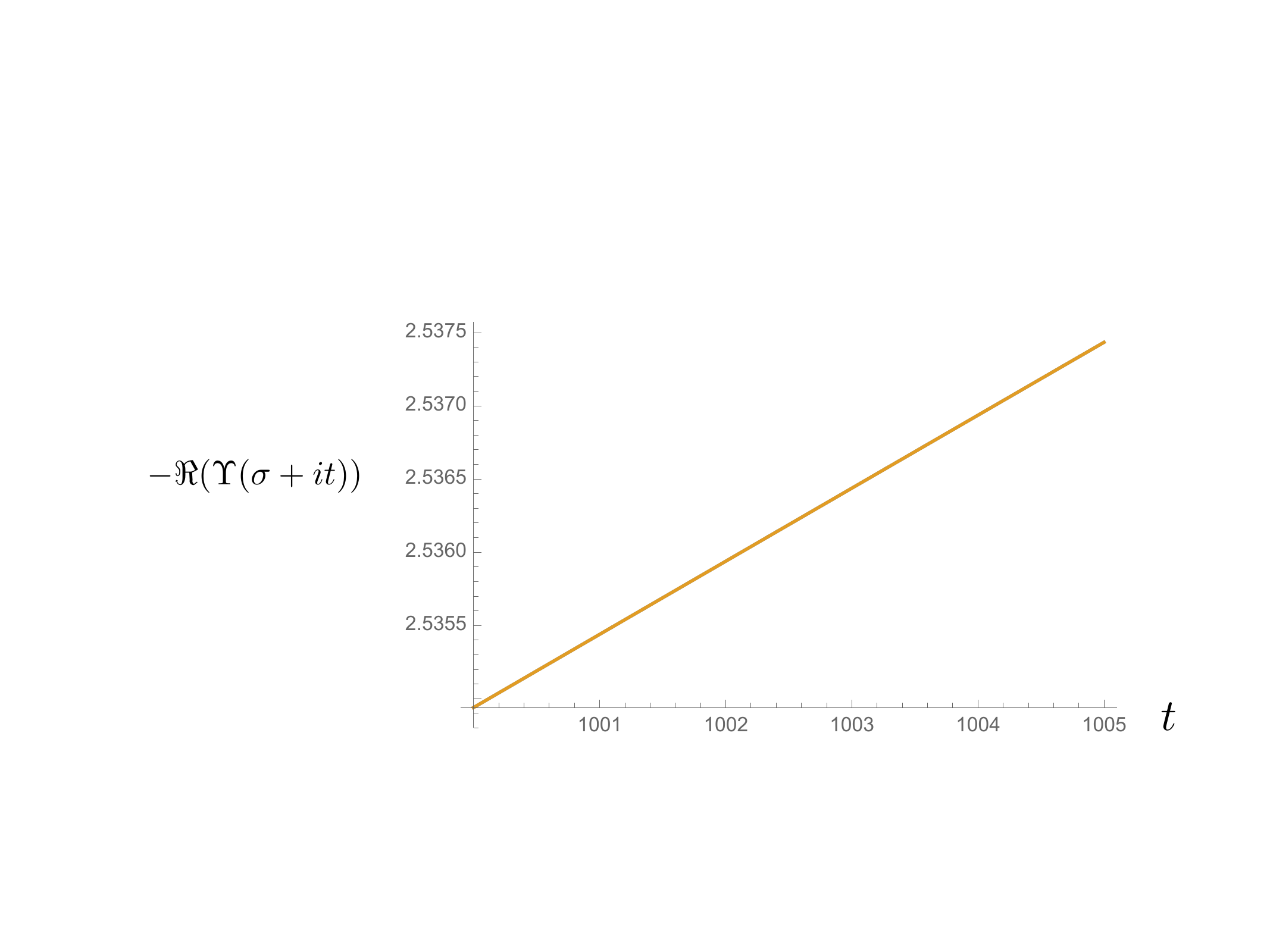}
\caption{A plot $-\Re (\ZpZ (s))$ on the critical line $s=\half  + i t$ and $\half \log(t/2 \pi)$  as a function of  $t$ for $1000< t < 1005$.   
The zeros are points on this line which are invisible at this resolution.}
 \label{PropHalf1000}
\end{figure}

\subsection{Analytic continuation of the Euler product for $\ZpZ (s)$.}

The Euler product has been a crucial ingredient in the above considerations.   
  Compared to $\log \zeta$ and $\arg \,\zeta$,  it is easier to analytically continue $\ZpZ$  into the strip and  it is better behaved there.      For this subsection let us indicate one analytical continuation of $\ZpZ (s)$ into the strip based on the prime-zeta function.    
 We will not use this result in the sequel,   rather presenting  it is meant to show that $\ZpZ$,    originally defined over primes \eqref{ZpZprimes}, 
 can be analytically continued in and of itself into the critical strip.    Using the Euler product,  one has 
\beq
\label{ZpZprimes}
- \ZpZ (s) = \sum_\pprime  \frac{\log \pprime }{\pprime^s -1},  ~~~~~\Re (s) > 1.
\eeq
Define the prime-zeta function $P(s)$:
\beq
\label{PrimeZeta}
P(s) = \sum_\pprime   \inv{\pprime^s} .
\eeq
Then expanding $1/(\pprime^s -1)$ one obtains the known formula 
\beq
\label{PrimeZeta2}
\ZpZ (s) = \sum_{n=1}^\infty P'(n s),
\eeq
for $\Re (s) > 1$.      Whereas the sum over primes in \eqref{ZpZprimes} does not converge,  it is known how to analytically continue $P(s)$ into the 
critical strip,   and the sum over $n$ in \eqref{PrimeZeta2} converges.

 \section{Spectral flow for a non-hermitian example which violates the RH}

In order to sharpen the above ideas,   it is very useful to have an example which has almost all of the properties of $\zeta (s)$ but the RH is violated.
There is a well-known such example due to Davenport-Heilbronn \cite{DavenportH}.   It satisfies a functional equation like $\zeta (s)$,  it has an infinite number of zeros on the line and also off the line.    As we explain,   the main difference is that it does not have an Euler product representation.
Recall that for $\zeta (s)$ the Euler product was necessary to define a unitary S-matrix.   
Without the Euler product,  one cannot even define the quantum scattering problem as in Section II \cite{LecMussDefect}. 
Without a unitary S-matrix,   the underlying hamiltonian must be non-hermitian,  and this implies that there should be complex eigenvalues which violate the RH.    In this section we study the spectral flow for the eigenvalues $E_n (\sigma)$ and show that indeed the analog of Proposition 
\ref{Prop1} is violated.

\def\rhozero{\rho_{\bullet}}
\def\rhozeroConj{\bar{\rhozero}}
\def\tzero{t_\bullet}

\def\ZpZDH{\ZpZ}

\bigskip

Let $L(s, \character)$ denote the Dirichlet $L$-function based on the non-principal and primitive mod $q=5$  character 
\beq
\label{chis}
\{\character(1), \character(2), \ldots, \character (5)\} = \{ 1, i, -i, -1, 0 \},       ~~~~~(q=5). 
\eeq
Due to the complete  multiplicity of the characters, $X(n) X(m) = X(nm)$,  it has both a series representation and an Euler product as in
\eqref{LDH}:
\beq
\label{LDH}
L(s, \character) = \sum_{n=1}^\infty  \frac{\character (n)}{n^s}    
 \, =\,\, \prod_{j=1}^{\infty} \(1 - \frac{\character(\p)}{\p_j^s}\)^{-1}  ~~~~~(\Re (s) > 1 ).
\eeq 
The DH function $\CD (s)$ is constructed  to satisfy a duality functional equation.  It is defined by the linear combination 
\beq
\label{DH}
\CD (s) \equiv  \tfrac{( 1- i \kappa)}{2} \, L(s,  \character ) +  \tfrac{( 1 + i \kappa)}{2} \, L(s,  \bar{\character}) 
\eeq
where $\kappa \equiv 
\tfrac{ \sqrt{10 - 2 \sqrt{5}} -2}{\sqrt{5} -1 }$.
Each $L(s, \character)$ function in the definition \eqref{DH}   has the   Euler product formula representation  in \eqref{LDH},   
however the linear combination in $\CD (s)$ does not. 

For the remainder of this section we use the same notation $\chi (s)$,   $\theta(\sigma, t)$, $\vartheta (\sigma,t)$,  $t_n$,  $E_n (\sigma)$,  and $\ZpZ$  as above for $\zeta$,  however they all refer to the
DH function.  
Define it's completion 
\beq
\label{xiDH}
\chi (s) \equiv \( \tfrac{\pi}{5}  \)^{-s/2} \Gamma \( \tfrac{1+s}{2} \) \, \CD (s),
\eeq
which satisfies 
$\chi (s) = \chi (1-s)$ by design.     
As for $\zeta (s)$ we consider its  argument: 
\beq
\label{thetaDS} 
\theta (\sigma, t) = \arg \, \chi  (s) =  \thetaRS (\sigma, t)  +   \arg \, \CD (s); ~~~~~\thetaRS  (\sigma,t)) = \arg\,  \Gamma \(  \tfrac{s+1}{2}  \) - \tfrac{t}{2} \log ( \tfrac{\pi}{5}) . 
\eeq
The appropriate Riemann-Siegel $\thetaRS$  on the critical line is
$\thetaRS (t) \equiv \thetaRS(\half, t)$.  
The analog of eqn.  \eqref{Horzn}  for the hypothetical $n$-th zero $\rho = \half + i t_n$ of the DH function on the critical line is simply \cite{FrancaLeClair}: 
\beq
\label{transDH}
\lim_{\delta \to 0^+} \theta(\half+\delta, t_n) =  \( n- \tfrac{1}{2} \) \pi .
\eeq
The counting function for the number of zeros up to height $T$ in the {\it entire} critical strip as usual follows from the argument principle: 
\beq
\label{NoftDH} 
N(T) = \lim_{\delta \to 0^+}  \theta(\half  + \delta, T)/\pi.
\eeq

As for $\zeta (s)$  we deform  the equation to obtain equations for $E_n (\sigma)$ where 
$t_n = \lim_{\sigma \to \half^+}  E_n (\sigma)$.   (For more details see \cite{LecMussDefect}.)    
In the limit of large  enough $E_n$ 
\beq
\label{FLDaveAss}
\frac{E_n (\sigma) }{2} \log \( \frac{q \, E_n (\sigma) }{2 \pi e} \)  + \arg \, \CD (\sigma + i E_n  (\sigma) ) = (n- \half) \pi.
\eeq
The sectral flow equation is of the form \eqref{dEn}.
For this case,   $\tstar \gtrsim 3$.   
Repeating arguments from Section III,   the RH for this $L$-function would be true if the following condition is satisfied
\beq
\label{DHProp}
- \Re \( \ZpZ (s) \) < \inv{2} \log \( \frac{5t}{2 \pi} \),  ~~~{\rm for} ~~ \Re (s) > \half,  ~~~~\ZpZDH  (s) \equiv  \frac{\CD' (s)}{\CD (s)}.
\eeq

\begin{figure}[b]
\centering\includegraphics[width=.6\textwidth]{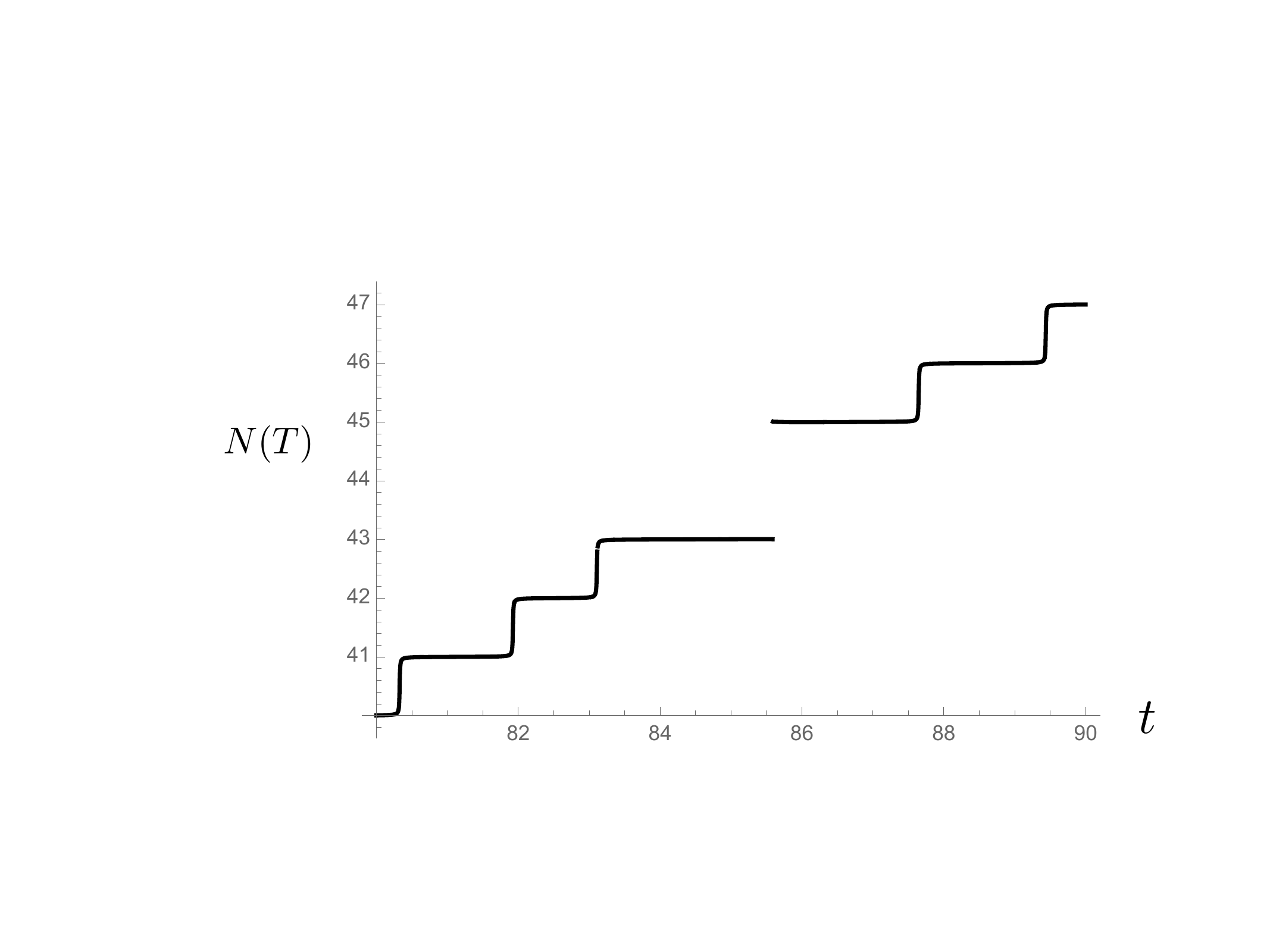}
\caption{A plot of $N(t)$ in \eqref{NoftDH} as a function of $t$ in the range of the first zero off the line.  We emphasize that 
the non-zero $\delta$ in \eqref{NoftDH} is necessary otherwise $N(t)$ would jump {\it discontinuously} by $1$ at zeros on the line.}
 \label{DHNofT}
\end{figure}

For illustration,   let us limit ourselves to the range $0<t<100$.   
One can easily check numerically that \eqref{transDH} correctly gives the first $43$ zeros,  which are all on the critical line.   
However for $n = 44$ and $45$ there is no solution to the equation \eqref{transDH}  \cite{FrancaLeClair,LecMussDefect}.
This can be seen from Figure \ref{DHNofT},   and   signifies two zeros off the line which are complex conjugates. 
They are: 
\beq
\label{DHoffLine} 
\CD (\rhozero) = 0 ~~~{\rm for} ~~ \rhozero = 0.8085171825 ~\pm ~ i \, 85.6993484854.
\eeq
There are an infinite number of zeros on the line which satisfy \eqref{transDH},   and also an infinite number of zeros off the line where there are no solutions to \eqref{transDH}.  
The next $n$ where the latter occurs is $n=63$.

\

\begin{figure}[t]
\centering\includegraphics[width=.7\textwidth]{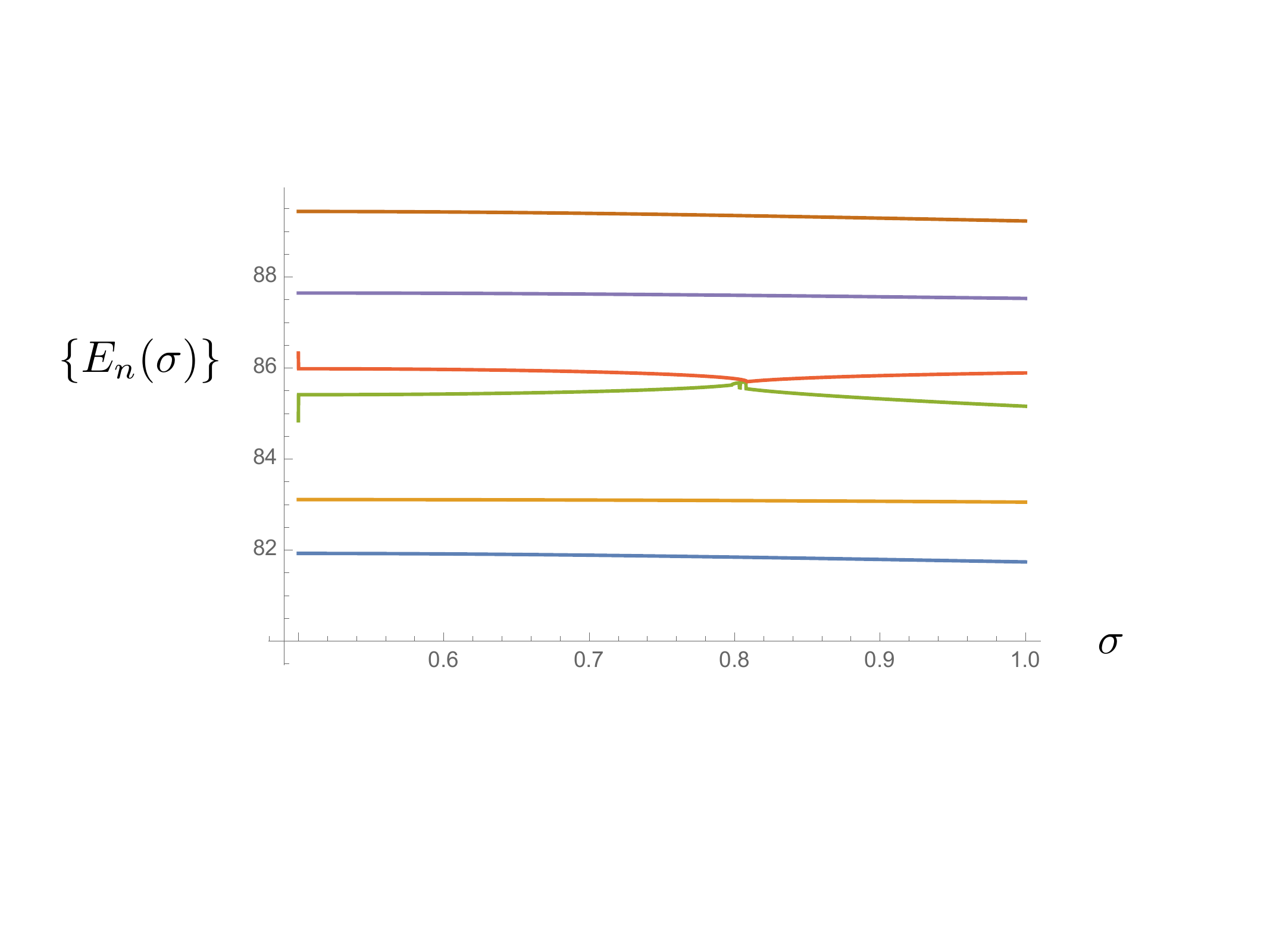}
\caption{A plot of $E_n (\sigma )$ as a function of $\sigma$  for $n=42, 43, 44, 45, 46, 47$  for the Davenport-Heilbronn function. One sees
that $E_n (\sigma)$ is ill-defined as $\sigma \to \half$ for $n=44,45$.}
 \label{LevelCrossing}
\end{figure}

\begin{figure}[t]
\centering\includegraphics[width=.5\textwidth]{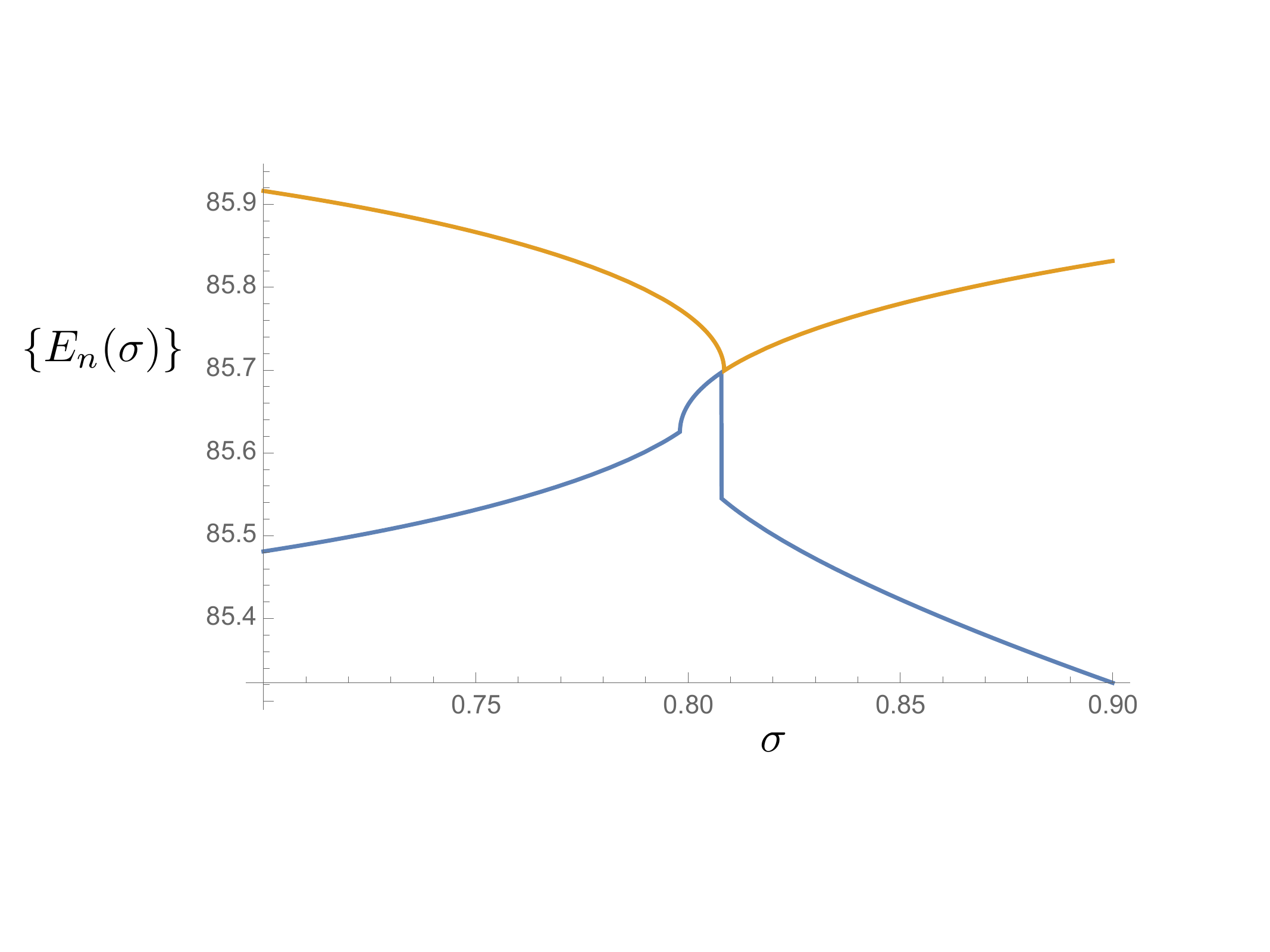}
\caption{A plot of $E_n (\sigma )$  as a function of $\sigma$ for $n=44, 45$  for the Davenport-Heilbronn function. The zeros off the critical  line occur at 
$\rho_\bullet  = 0.8085171825 ~\pm ~ i \, 85.6993484854$.}
 \label{CloserLook}
\end{figure}

Let us now see how this is consistent with the spectral flow condition \eqref{DHProp},   namely specifically how it is violated.  
In Figure \ref{LevelCrossing} we plot $E_n (\sigma)$ as a function of $\sigma$ for $n$ in the vicinity of zeros off the line.  
Near the zero off the line,   one observes what was anticipated in the Introduction,   namely a pair of eigenvalues $E_n$ which are well-defined 
for $\sigma \gtrsim  0.81$  coalesce into a pair of complex eigenvalues.     Closer inspection in Figure \ref{CloserLook} suggests a level-crossing.
In any case $dE_n/d\sigma $ is infinite there.   
Furthermore,   the $E_n (\sigma)$ are not well defined   as $\sigma \to \half^+$,  as can also be seen in the Figure \ref{LevelCrossing}.

\begin{figure}[t]
\centering\includegraphics[width=.6\textwidth]{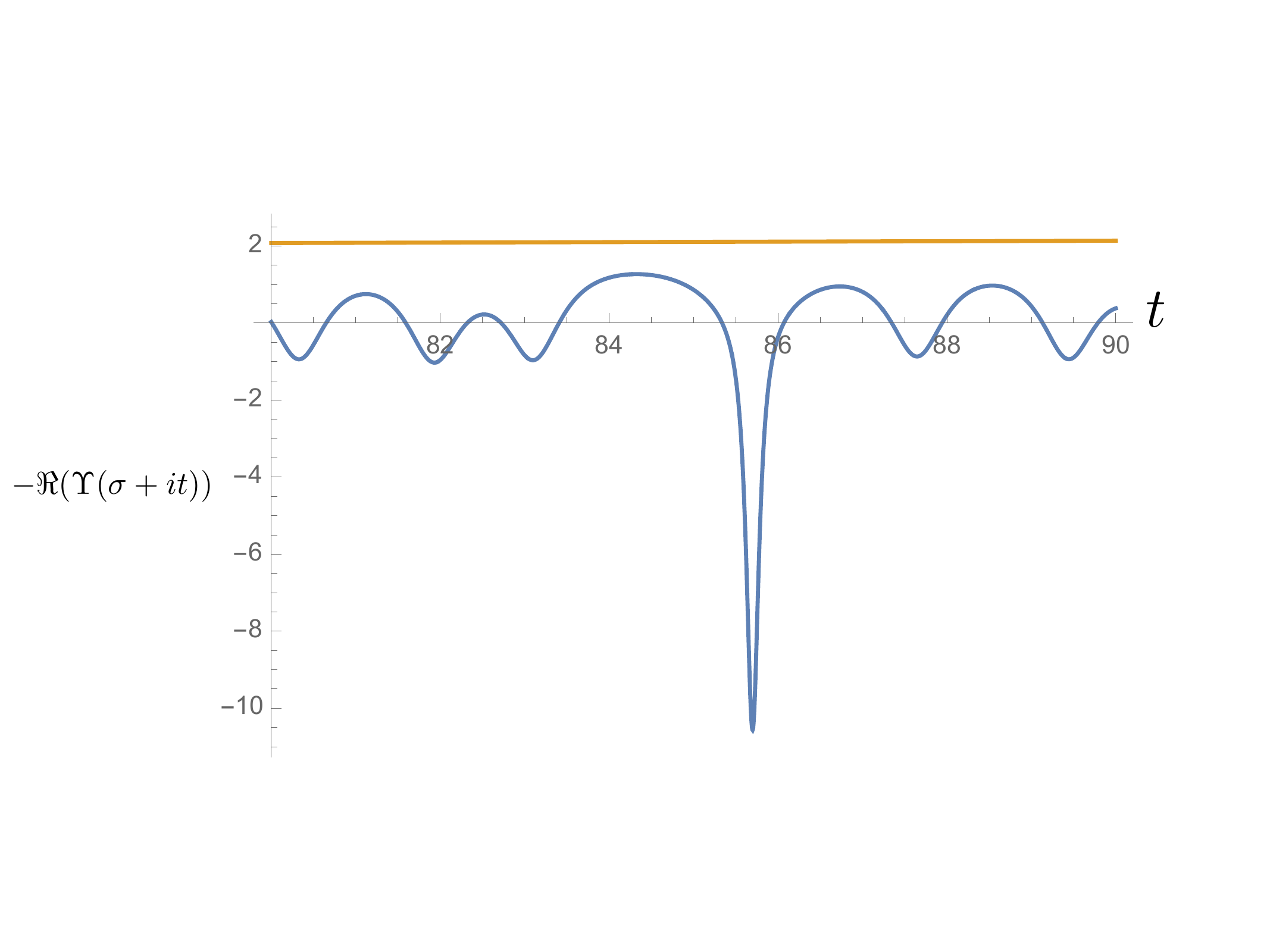}
\caption{A plot of $-\Re(\ZpZDH (\sigma + i t)$ verses $\half \log (5 t/2\pi)$ 
 for the Davenport-Heilbronn function as a function of $t$ for $\sigma=0.9$.}
 \label{DH1}
\end{figure}

\begin{figure}[t]
\centering\includegraphics[width=.7\textwidth]{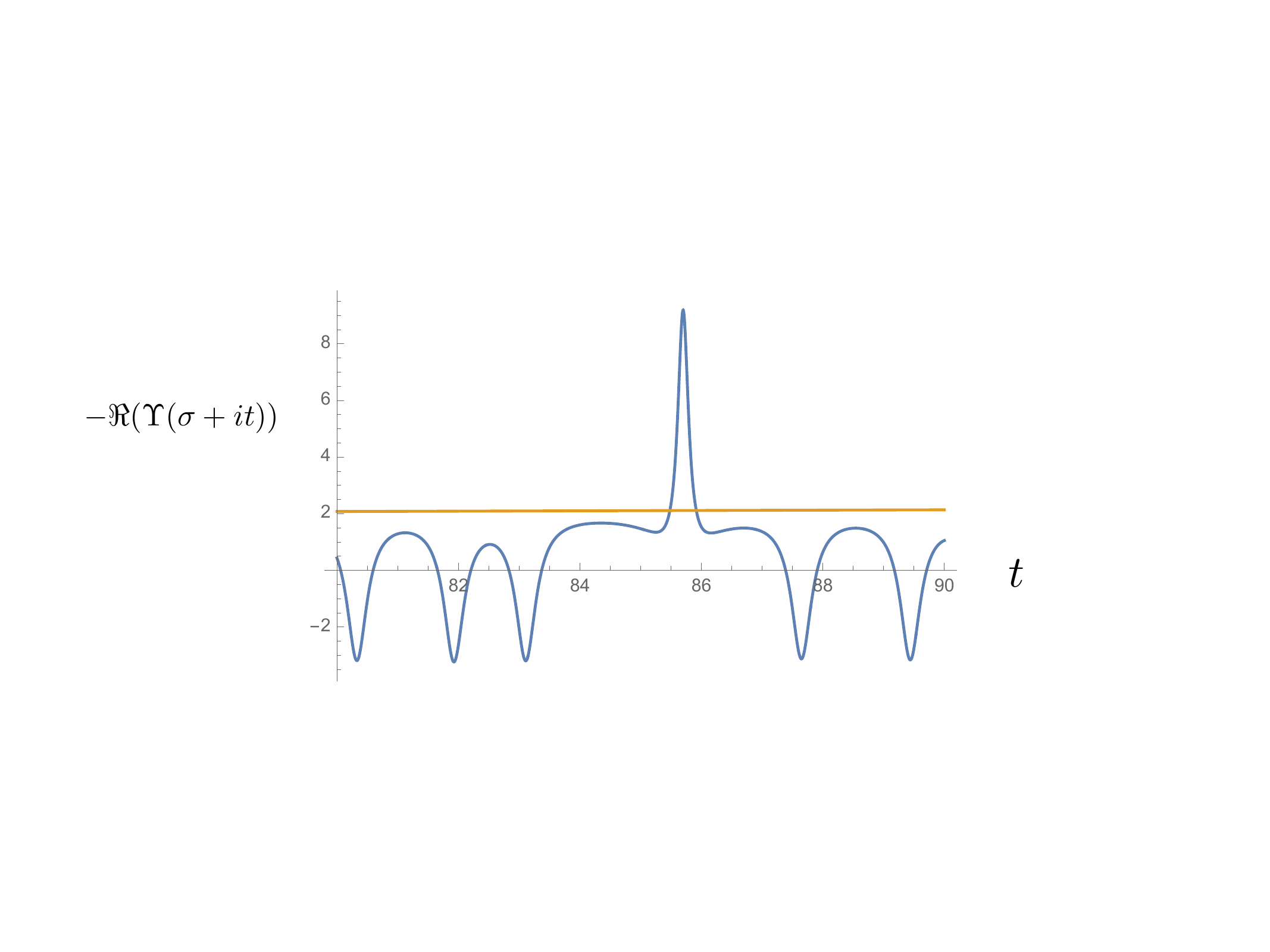}
\caption{A plot of $-\Re(\ZpZDH (\sigma + i t)$  
 for the Davenport-Heilbronn function verses $\half \log (5 t/2\pi)$ as a function of $t$ for $\sigma=0.7$.}
 \label{DH2}
\end{figure}

In the range $0<t<100$ there are no zeros off the line with $\sigma > 0.9$,   thus \eqref{DHProp} should be valid,  and it is as can been seen 
from Figure \ref{DH1}.    On the other hand,   for $\sigma > 0.7$ the condition \eqref{DHProp} should be violated due to the zero in
\eqref{DHoffLine},  and indeed it  clearly is,  as can be seen in
Figure \ref{DH2}.

\section{Analytic proof of the Propositions for $\zeta (s)$ assuming the RH}

Thus far we have argued for Propositions \ref{Prop1}, \ref{Prop2} without assuming the RH,    namely it would follow from the hermiticity of the underlying hamiltonian of the LM model due to its unitary S-matrix
and Bethe ansatz equation,  as stated in Proposition \ref{Prop0}.    To emphasize the logic,   the RH is true if the LM model defines a hermitian eigenvalue problem.       
In this section we show that the latter statement and  the Propositions of Section III  can be derived if one 
{\it assumes}  the RH,   {\it which we did not above}.    Thus,    the result of the following analysis is already a consequence of the above Propositions.            

Lagarias proved \cite{Lagarias} that the RH is true if 
\beq
\label{Lagarias}  
\Re \(  \frac{\xi'(s)}{\xi (s)} \) > 0~~~~~{\rm for} ~ \Re(s) > \half, ~~~~~ \xi(s) \equiv 
s (s-1)\, \chi (s) /2.
\eeq
Using known results found  in  \cite{DavenportBook},  it can be shown that this implies Proposition \ref{Prop1}. 
Let us show this directly in a somewhat simpler fashion by working directly with $\zeta (s)$,   although this argument must be implicit in Lagarias'
work.         
By taking a logarithmic derivative of the functional equation $\xi(s) = \xi(1-s)$ and the Hadamard product,  
and also utilizing the Euler product,   one can show \cite{DavenportBook}
\beq
\label{DavenportEq}
- \ZpZ (s)  = \inv{s-1} - B - \log \sqrt{\pi}  + \inv{2}  \frac{ \Gamma' (s/2 +1) }{\Gamma(s/2 +1)}   
-\sum_\rho \( \inv{s-\rho} + \inv{\rho} \),
\eeq
where
\beq
\label{Bdef}
B = - \gamma_E /2 -1 + \log( 2 \sqrt{\pi} ) = -0.0230957...
\eeq
with $\gamma_E$  the Euler constant.  
Above,   $\sum_\rho$  denotes a sum over all non-trivial zeros inside the critical strip $0 < \Re (\sigma) < 1$,  including potential zeros off the critical line.  
Using the Stirling formula,
\beq
\label{Dave2}
- \Re ( \ZpZ (s) ) = 
 \half \log (t/2\pi) \,+ \,
 \Re\( \inv{s-1} \) -B + \frac{(2 + 6 \sigma + 3 \sigma^2 )}{12 t^2}  
-\Re \, \sum_\rho \( \inv{s-\rho} + \inv{\rho} \) + \CO(1/t^4).
\eeq

Let us turn to $\sum_\rho$.     Given  a zero at  $\rhozero = \sigmazero + i \tzero$,  the functional equation and complex conjugation implies 
there are  a total of 4 zeros at $\rhozero = 1- \sigmazero \pm  i \tzero$ and $\rhozero = \sigmazero \pm i \tzero$.      
Now  $\Re (1/\rhozero) = \sigmazero/|\rhozero|^2$.     Thus in the sum over $\rhozero$,   zeros off the line $\sigmazero \neq \half$ cancel
and one is left with a sum over zeros on the critical line:
\beq
\label{sum1}
\sum_\rho \, \Re \( \inv{\rho} \) =   \sum_n  \inv{|\rho_n|^2}, ~~~~~ \rho_n = \half + i t_n,  ~~~~ t_n > 0
\eeq
The above sum actually converges and cancels the $B$ term in \eqref{Dave2}.     For instance,   for the first 200,000 zeros we find   
\beq
\label{sumzeros}
 \sum_{n=1}^{2 \cdot 10^5} \,   \inv{|\rho_n|^2} = 0.0230832. 
\eeq
For $\half < \sigma < 1 $,  one sees that  $\Re (1/(s-1) )< 0$.    
Finally one has
\beq
\label{sum2}
\Re \( \inv{s - \rhozero} \) = \frac{\sigma - \sigmazero}{(\sigma - \sigmazero)^2 + (t-\tzero)^2} .
\eeq
If there is a zero off the line with $\sigmazero >\half$ then for $\sigma < \sigmazero$ the above term is negative and this spoils the bound
\eqref{conjecture2}.    
Assuming the RH,  all $\sigmazero = \half$ and the above is positive for $\sigma > \half$,    thus  the bound \eqref{conjecture2} is satisfied.    
This proves Proposition \ref{Prop1}  in the form \eqref{conjecture2}.     
 
Again assuming the RH,   on the critical line the expression \eqref{sum2} vanishes.     Neglecting all $1/t^2$ terms,  including those 
from $\Re (1/(s-1) )$,   which is valid for large $t$,   one obtains  Proposition \ref{Prop2}.

\section{Generalized and Grand Riemann Hypothesis}

The Riemann zeta function is a special case of an infinite class of $L$-functions based on Dirichlet characters;   it corresponds to that of the principal character mod $q=1$.       It has been conjectured that the RH is true for all such functions,  and this is referred to as the Generalized Riemann Hypothesis.        These functions enjoy both a functional equation relating $s$ to $1-s$,  and an Euler product representation.   
A completely different class of $L$-functions are built upon the Fourier coefficients of modular forms,    and also satisfy a functional equation and Euler product.    The conjecture that they also satisfy a RH is referred to as the Grand Riemann Hypothesis.   
The random walk argument for these classes of functions was presented in \cite{LMDirichlet,MLRW,FL2}.    Here we focus on arguments based on the LM model,  which are easily generalized to these two infinite classes of $L$-functions.  

The above spectral flow equations are easily generalized with little detailed analysis to these classes of functions,  thus we present the analogs of the 
above Propositions to these cases.     Because of the existence of the Euler product,   one can easily define the impurity scattering problem as we did for $\zeta (s)$ for a finite number $N$ of impurities,  define the quantized energies $E_n (\sigma)$,   and take the thermodynamic limit.    One thereby obtains the equations for individual zeros on the 
critical line $\Re (s) = \half$ proposed in \cite{FrancaLeClair} in the limit $\sigma \to \half$.           Based on such equations it is straightforward to obtain simple criteria for the RH to hold based on the spectral flow argument of Section III.        As above for Riemann $\zeta$,   we drop $\CO(1/t^2)$ corrections to the dispersion relation coming from the analog of Riemann-Siegel 
$\thetaRS$,  since they are negligible for large enough $t$,  which in fact is not so large,  again around $t \gtrsim 10$.   Below the statement
``for $t$ large enough'' refers to this.

\subsection{Generalized Riemann Hypothesis}

Let $L(s, \character)$ denote the $L$-function based on a primitive Dirichlet character $\character$ of modulus $q$.
The characters $\character$ are periodic,  $\character (n) = \character (n+q)$,   and are all phases which are roots of unity,  $\character (n) = e^{i \charphi (n)}$.  Thus these $L$-functions are already incorporated in the general LM model \eqref{Sjs}.    
Due to the complete multiplicity of the characters $X(n) X(m) = X(nm)$,    it has both a series and a Euler product representation:
\beq
\label{LDH2}
L(s, \character) = \sum_{n=1}^\infty  \frac{\character (n)}{n^s}    
 \, =\,\, \prod_{j=1}^{\infty} \(1 - \frac{\character(\p)}{\p_j^s}\)^{-1}  ~~~~~(\Re (s) > 1 ).
\eeq
The equations in \cite{FrancaLeClair}  for individual zeros on the line depend on various details such as the Gauss sum and other number-theoretic quantities,  but for our purposes these are negligible for large enough $t$.\footnote{See for instance equation (43) in \cite{FrancaLeClair}.}
Repeating the arguments of Section III we can propose the following. 

\bigskip
\begin{proposition}
The Generalized Riemann Hypothesis is true if for $t > \tstar$ large enough 
\beq
\label{GenRH}
- \Re \( \ZpZ (s) \)  <  \inv{2} \log \( \frac{q t}{2 \pi} \),  ~~~~ 
\ZpZ (s) \equiv \frac{ L'(s,\character )}{L(s, \character)} 
\eeq
for $\Re (s) > \half$.    
\end{proposition}
\bigskip
Here $\tstar$ is  typically smaller than for $\zeta$.      

\subsection{Grand Riemann Hypothesis}

For a detailed exposition of the mathematical background for  the $L$-functions in this section, we refer to Apostol's book \cite{Apostol}. 
Let $f(\tau)$ denote a cusp  form of weight $k$.     By definition under an $SL(2, \ZInteger)$ tranformation:
\beq
\label{SL2Z}
f \(  \frac{a \tau + b}{c \tau + d} \) = ( c \tau + d )^k \, f(\tau).
\eeq
One $SL(2, \ZInteger)$ tranformation implies the periodicity $\tau \to \tau + 1$,   
thus it has a Fourier series
\beq
\label{FourierTau}
f(\tau) = \sum_{n=0}^\infty a_f (n) \, q^n, ~~~~~q=e^{2 \pi i \tau} .
\eeq
From these Fourier coefficients one defines the $L$-function
\beq
\label{Ltau}
L_f (s) = \sum_{n=1}^\infty \frac{a_f (n)}{n^s} .
\eeq

It is well-known that it satisfies a functional equation and Euler product formula \cite{DavenportBook},  and 
the critical line is $\Re (s) = k/2$.   The Euler product is slightly more complicated since the coefficients 
$a_f (n)$ are not {\it completely} multiplicative,   but this does not affect our reasoning.   
  In \cite{FrancaLeClair} the analog of exact equations satisfied by individual zeros were presented, 
and this is all we need for our purposes:    
\beq
\label{TransEq}
t_n \log \( \frac{t_n}{2 \pi e} \) +  \lim_{\delta \to 0^+} \arg \, L_f (\tfrac{k}{2} + \delta + i t_n )  = \(n - \frac{ k + (-1)^{k/2} }{4} \) \pi .
\eeq
This leads to the Proposition:

\bigskip

\begin{proposition}
\label{Grand}
The Grand  Riemann Hypothesis is true if for $t$ large enough, $t > \tstar \approx 25$, 
\beq
\label{GrandRH}
- \Re \( \ZpZ (s) \)  <  \log \( \frac{t}{2 \pi} \),    ~~~~~ \ZpZ (s) \equiv  \frac{L'_f (s)}{L_f (s)} 
\eeq
for $\Re (s) > k/2$.  
\end{proposition} 

Note there is no $k$ dependence of the above bound,   and it differs from the $\zeta (s)$  case only in the overall coefficient of the $\log$ which is
$1$ rather than $\half$.    The statistical mechanical approach involving random walks was described in \cite{FL2}.

\begin{figure}[t]
\centering\includegraphics[width=.6\textwidth]{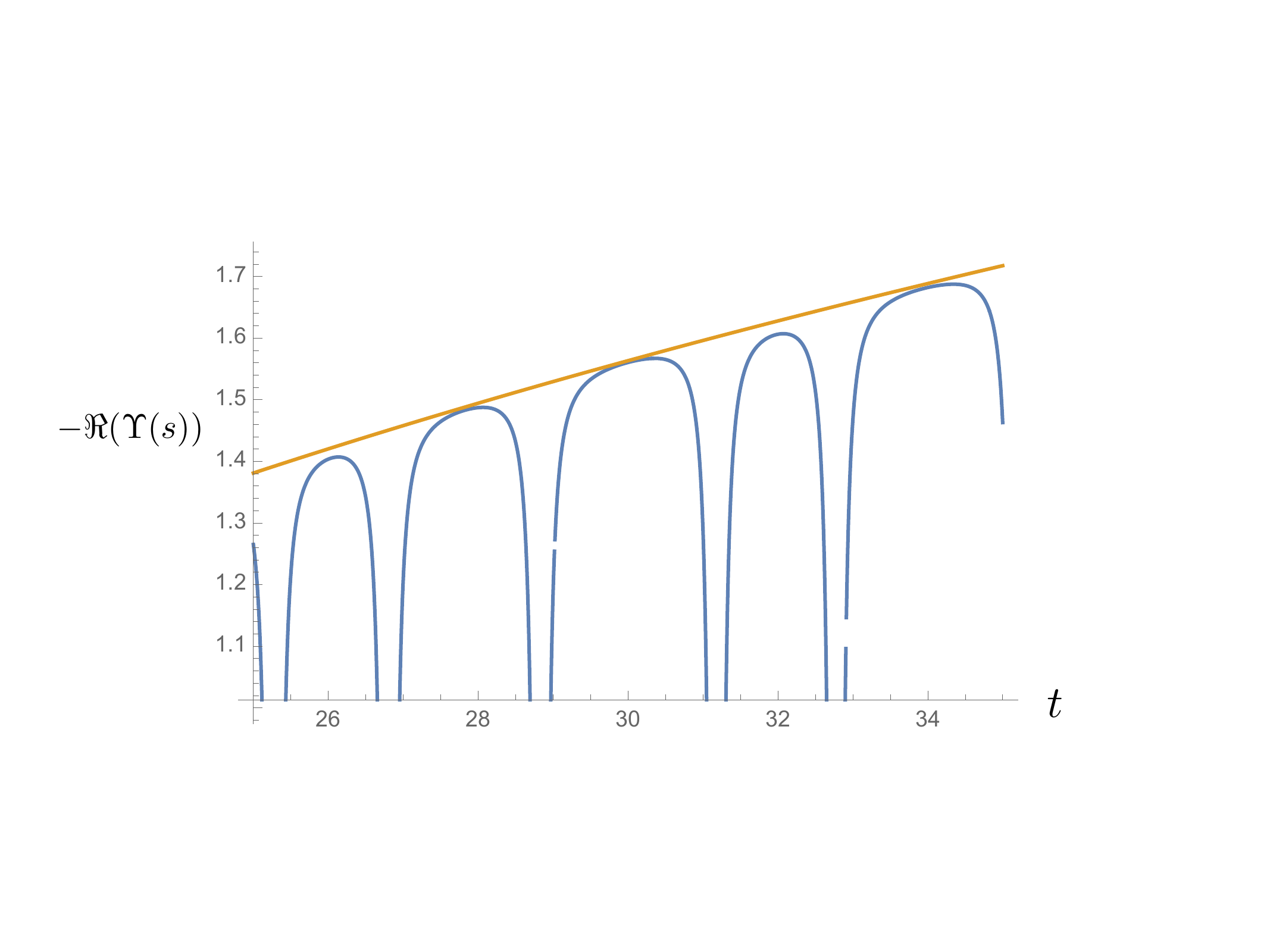}
\caption{A plot of $-\Re(\ZpZ (s))$ for the  weight $k=12$ modular $L_f$ based on Ramanujan $\tau$  
  verses $ \log (t/2\pi)$ as a function of $t$ for $\sigma$ just above the critical line,  $\sigma=6.01$.}
 \label{Ramanujan}
\end{figure}

The simplest example is the  modular form based  based on the Dedekind $\eta$ function: 
\beq
\label{etaFunction}
\eta (\tau) =  q^{1/24}  \prod_{n=1}^\infty ( 1- q^n ).
\eeq
The modular discriminant $\eta(\tau)^{24}$ is a weight $k=12$ modular form,   which is closely related to the inverse of the 
bosonic string partition function.\footnote{The coefficients $a_f (n)$ are known as Ramanujan ${\rm`` tau"}  (n)$}     
The $L_f$ function is computationally intensive even  for  not so large $t$.      Nevertheless we can provide numerical evidence for
Proposition \ref{Grand} in a region  $t> t_\star$,  with $t_\star  \approx 25$,  as shown in Figure \ref{Ramanujan}.

\section{Statistics for eigenvalues $E_n (\sigma)$ off the critical line.}

The Montgomery-Odlyzko conjecture is that the properly normalized  pair correlation function satisfies GUE statistics. 
In this section we compare the statistics of the actual zeros $t_n$ with that of $E_n (\sigma)$ for $\sigma>\half$. 
We carry out this comparison mainly out of curiosity,   and don't present any rigorous results as far as what this entails.

\def\deltaEk{\delta E_n^{(k)}}

Consider the normalized spacings
\beq
\label{deltatk}
\deltaEk (\sigma)   = \frac{1}{ 2\pi} \( E_{n+k}(\sigma)   - E_n (\sigma)  \) \log \(E_n (\sigma)/2\pi \) , ~~~ k=1,2, \ldots
\eeq
where $k=1$ corresponds to nearest neighbors, etc.  
The original conjecture concerns all pairs $k \geq  1$ and has the well-known distribution: 
 \beq
 \label{GUE}
 1- \frac{\sin^2 (\pi v)}{\pi^2 v^2 } = \sum_{k=1}^\infty \rho_k (v), 
 \eeq
 where the  function $\rho_1 (v)$ gives the nearest neighbor correlation and so forth.   
 The distribution $\rho_1$ is known as the Gaudin distribution,  which can  be  expressed in terms of Fredholm determinants,  however  has no simple expression.
It is known that a very good approximation to $\rho_1 (v)$ is the Wigner surmise:  
 \beq
 \label{Wigner}
 \rho _1 (v) \approx \rhoWigner (v) = \frac{32\, v^2 }{\pi^2}  \,e^{-4 v^2/\pi} .
 \eeq
  The above just equals the Maxwell-Boltzmann distribution of speeds $v$ in a free gas of non-relativistic particles of mass $m$, with kinetic energy $\tfrac{1}{2} m v^2 $,   
 at a temperature $k_B T = m\pi/8$.   

\begin{figure}[t]
\centering\includegraphics[width=.4\textwidth]{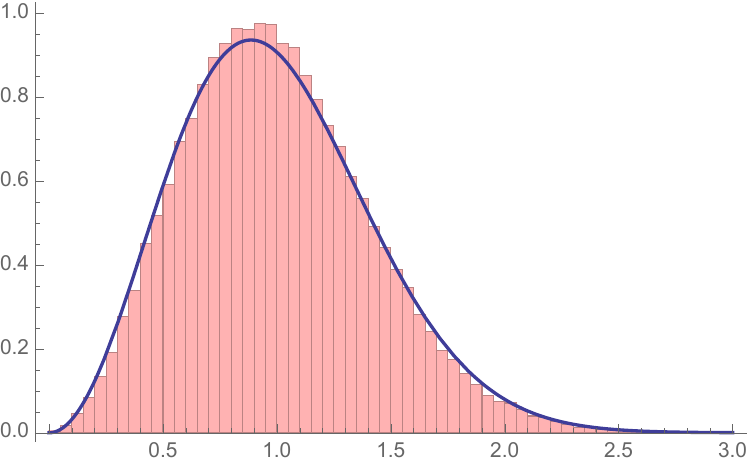}
\caption{Probability distribution for normalized consecutive eigenvalues $E_ n (\sigma)$ for $\sigma = \half$ in the range $10^6 < n <10^6 + 10^5$.  The solid curve is the 
Wigner surmise.}
 \label{GUE1}
\end{figure}

\begin{figure}[t]
\centering\includegraphics[width=.4\textwidth]{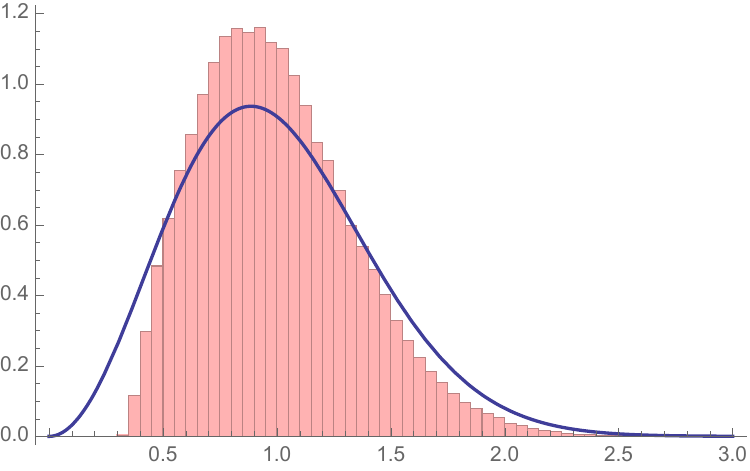}
\caption{Probability distribution for normalized consecutive eigenvalues $E_ n (\sigma)$ for $\sigma = 0.6$ 
 in the range $10^6 < n <10^6 + 10^5$.     The solid curve is the 
Wigner surmise.}
 \label{GUE2}
\end{figure}

Since we only wish to compare the statistics of the true zeros  $t_n$  with those of $E_n (\sigma)$ for $\sigma > \half$,   for simplicity   we only consider the  nearest neighbor pair correlation $\rho_1 (v)$ and compare it with the Wigner surmise since this requires less computational effort.  
In Figure \ref{GUE1} we present  a normalized histogram for $\rho_1$ for $10^5$ eigenvalues $E_n (\sigma)$ for $\sigma = \half$ which are the actual zeros $t_n$.
The results are what we expect,  namely that the results are consistent with random matrix theory.  
 In Figure \ref{GUE2} we present analogous results for $\sigma = 0.6$.    One can readily conclude that the $E_n (\sigma)$  are not consistent with GUE statistics except for $\sigma = \half$.    In particular,  above $\sigma = \half$ there appears to be stronger level-repulsion since the probability density is zero in a finite range above $0$ in comparison with the Wigner surmise.

\section{Conclusions}

We proposed that the quantized energies $\{ E_n (\sigma) \}$ for the LM model are real due to the fact that its S-matrix is unitary based
on the Euler product  and the hamiltonian for the free part,  i.e. the dispersion relation,   is hermitian (Proposition \ref{Prop0}.)      Without assuming the RH,     we studied the spectral flow, 
 namely how
the $E_n (\sigma)$ depend on $\sigma$ and derived a simple condition for all eigenvalues to be real,  stated as Propositions \ref{Prop1} and  
\ref{Prop2} above.      We showed how these Propositions are 
equivalent to the validity of the Riemann Hypothesis.       We extended this reasoning to the two infinite classes of functions for 
the Generalized and Grand Riemann Hypotheses.     We also presented spectral flow arguments for a counterexample where the 
RH is violated,  and attributed this to the lack of an Euler product which implies a non-unitary S-matrix.   
This work provides ample evidence that for the RH to be true,  one needs both the functional equation and the Euler product.  

Based on this article,   the validity of the RH  comes  down  Proposition \ref{Prop0},   namely that  the  hamiltonian for the LM model and its generalizations has real eigenvalues $E_n (\sigma)$.  
We argued that this is valid due to the unitarity of the S-matrix and  the reality and invertibility of the dispersion relation.     
In fact, it is clear from \eqref{EofpSmall} that the free hamiltonian for our dispersion relation is a hermitian operator.    
We thus cannot see any obstructions to  Proposition \ref{Prop0} at present.    
This work shows that the Hilbert-P\'olya conjecture can be  true if one plucks a Hamiltonian of  
the right feather, in particular one that leads to a scattering problem rather than a bound state problem.
We point out that 
 the recent work \cite{Negro} seems relevant to the study of the free non-interacting part of the hamiltonian if the dispersion relation is based on the exact Riemann-Siegel $\thetaRS$ function.

An appealing feature of the above Propositions is that they all take a universal form,  namely  that the appropriate quantity,   $- \Re(\ZpZ (s) )$, 
 is bounded by simple 
variations of $\log (t / 2 \pi)$ for  both the Generalized and Grand Riemann Hypotheses,   and this suggests a simple,   universal strategy
towards  understanding all of them.   This is in contrast to most ideas based on Hilbert-P\'olya which are idiosyncratic to only $\zeta (s)$ 
itself and do not indicate how to proceed  beyond  to these Generalized and Grand Riemann Hypotheses.

There are many interesting avenues for further exploration.    For the $L$-functions that violate the RH,  such as the example in Section IV,   it would be interesting to study if the complex eigen-energies satisfy the statistics of random non-hermitian hamiltonians classified in \cite{BL38}.  

The LM model is more general than the specific realization we studied here where the S-matrix was defined using the prime numbers.  
However one can replace the primes with other real numbers and still have a well-defined scattering problem with impurities \cite{LecMussDefect}.  
If the impurities are random phases then this defines an interesting problem in Anderson localization.    For the version  of the LM model in this paper,  the randomness of the impurities stems from the pseudo-randomness of the primes.

\section{Acknowledgements} 

We  especially thank  Giuseppe Mussardo for our collaboration in \cite{LecMussDefect} which led to the considerations of this paper. 
We also thank the Simons Center for Geometry and Physics and the organizers  for their  invitation to their recent program in May 2024 on
far from equilibrium physics which stimulated some of these ideas,  in particular conversations with  J.  Verbaarschot.

\end{document}